%% file: main.tex
\documentclass[a4paper]{article}
\usepackage{color}
\usepackage{ifthen}
\usepackage{bmpsize}
\usepackage{amsmath}
\usepackage{amssymb}
\usepackage{amsfonts}
\usepackage{float}
\usepackage{array}
\usepackage[absolute]{textpos}
\usepackage{wrapfig}
\usepackage{multicol}
\usepackage{boxedminipage}
\usepackage{lettrine}
\usepackage{times}
\usepackage[intoc]{nomencl}
\usepackage{graphicx,xcolor,import}
\usepackage{caption}
\usepackage{subcaption}
\usepackage{footnote}
\usepackage[numbers]{natbib}
\usepackage{algorithm}
\usepackage{algpseudocode}
\usepackage{float}
\usepackage{todonotes}
\usepackage{authblk}

\DeclareMathOperator*{\argmax}{arg\,max}

 \usepackage{tikz}
\usepackage[a4paper,top=3cm,bottom=2cm,left=3cm,right=3cm,marginparwidth=1.75cm]{geometry}

\title{Autonomous Materials Discovery Driven by Gaussian Process Regression 
with Inhomogeneous Measurement Noise and Anisotropic Kernels}
\author[1,5]{Marcus Noack*}
\author[2]{Gregory S. Doerk}
\author[3]{Ruipeng Li}
\author[4]{Jason K. Streit}
\author[4]{Richard A. Vaia}
\author[2]{Kevin G. Yager}
\author[3]{Masafumi Fukuto}
\affil[1]{The Center for Advanced Mathematics for Energy Research 
Applications (CAMERA), Lawrence Berkeley National Laboratory, Berkeley, CA 94720}
\affil[2]{Center for Functional Nanomaterials, Brookhaven National Laboratory, Upton, NY 11973}
\affil[3]{National Synchrotron Light Source II, Brookhaven National Laboratory, Upton, NY 11973}
\affil[4]{Materials and Manufacturing Directorate, Air Force Research Laboratories, WPAFB, OH 45433}
\affil[5]{MarcusNoack@lbl.gov}

\begin{document}
\maketitle

\begin{abstract}
A majority of experimental disciplines face the challenge of exploring large and high-dimensional parameter spaces in search of new scientific discoveries. Materials science is no exception; the wide variety of synthesis, processing, and environmental conditions that influence material properties gives rise to particularly vast parameter spaces.
Recent advances have led to an increase in
efficiency of materials discovery by increasingly automating the exploration processes.
Methods for autonomous experimentation have become more 
sophisticated recently, allowing for multi-dimensional
parameter spaces to be explored efficiently and with minimal human
intervention, thereby liberating the scientists to focus 
on interpretations and big-picture decisions. 
Gaussian process regression (GPR) techniques have emerged as 
the method of choice for steering many classes of experiments. 
We have recently demonstrated the positive impact of GPR-driven  decision-making algorithms on autonomously 
steering experiments at a synchrotron beamline. However,
due to the complexity of the experiments, GPR often cannot be used
in its most basic form, but rather has to be tuned to account for
the special requirements of the experiments.
Two requirements seem to be of
particular importance, namely inhomogeneous measurement noise (input dependent or non-i.i.d.) 
and anisotropic kernel functions, 
which are the two concepts that we tackle in this paper. 
Our synthetic and experimental tests demonstrate the importance of both
concepts for experiments in materials science and the benefits that result
from including them in the autonomous decision-making process. 
\end{abstract}
\section{Introduction}
Artificial intelligence and machine learning 
are transforming many areas of experimental science.
While most techniques focus on analyzing ``big data'' 
sets, which are comprised of redundant information, 
collecting smaller but information-rich data sets
has become equally important. 
Brute-force data collection
leads to tremendous inefficiencies in the utilization of experimental facilities and instruments, in data analysis and
data storage; large experimental facilities around the globe are running
at 10 to 20 percent utilization and are still spending
millions of dollars each year to keep up with the increase in the amount of 
data storage needed \cite{habib2016ascr,gerber2018crosscut,almgren2017advanced,thayer2019data}. In addition, 
conventional experiments require scientists to prepare samples 
and directly control experiments, which leads to highly-trained researchers
spending significant effort on micromanaging experimental tasks rather than thinking about scientific meaning. 
To avoid this problem, autonomously steered experiments 
are emerging in many disciplines. These techniques place measurements only
where they can contribute optimally  to the overall knowledge
gain. Measurements that collect redundant information are avoided.
These autonomous approaches minimize the number of needed measurements to reach a certain model
confidence, thus optimizing the utilization of 
experimental, computing and data-storage facilities.

A universal goal in materials science is to explore
the characteristics of a given material 
across the set of all conceivable combinations of experimental parameters, which can be thought of as a parameter space defining that class of materials. 
The experimental parameters can be the characteristics of material components, their composition, processing or synthesis
parameters, and environmental conditions on which the experimental outcomes 
depend \cite{pilania2013accelerating,jain2013commentary}. 
Successful exploration of the parameter space amounts to being able to
define a high-confidence map, i.e. a surrogate model function,
of experimental outcomes across all elements of the set. For two-dimensional parameter spaces, this is traditionally achieved by ``scanning'' the space, often on a simple Cartesian grid. 
Selecting a scanning strategy implies picking a scan resolution without
knowing the model function. 
When the parameter
space is high-dimensional, 
an approach based on intuition is often used, i.e.,
manually
selecting measurements, assessing trends and patterns
in the data, and selecting follow-up measurements.
With increasing dimensionality of the parameter space, this method quickly fails
to efficiently explore the space and becomes prone to bias.
Needless to say, the human brain is generally poorly
equipped for high-dimensional pattern recognition.

What is needed are methods that decouple the human from
the measurement selection process. 
This fact served as a motivation to establish a research field called
design of experiment (DOE) \cite{dean2000design}, which can be traced back as far as the late 1800s.
These DOE methods are largely geometrical, independent 
of the measurement outcomes  
and are concerned with efficiently exploring the entire parameter space. 
The latin-hyper-cube method is the 
prime example of this class of methods
\cite{mckay1979comparison,fisher1992arrangement}.
Most of the recent approaches to steer experiments 
are part of a field called active learning, which 
is based on machine learning techniques 
\cite{pilania2013accelerating,scarborough2017dynamic,
godaliyadda2016supervised,balachandran2016adaptive}.
Others have used deep neural networks to make data acquisition cheaper \cite{cang2018improving}.
Many techniques originated from image analysis 
\cite{godaliyadda2016supervised,martinez2007image},
but, as images are traditionally two or three dimensional,
these methods rarely scale efficiently to high-dimensional
spaces.
A useful collection of methods can be found 
in \citet{santner2003design} and \citet{forrester2008engineering}.

Gaussian process regression (GPR) is a particularly successful technique to steer experiments autonomously \cite{noack2019kriging,hanuka2019online}. 
The success of GPR in 
steering experiments is due to its non-parametric nature; simply speaking,
the more data that is gathered the more complicated the model 
function can become. The
number of parameters of the function, 
and therefore its complexity, does not have to be
defined a priori.
This is in contrast to neural networks, which need a specification of an
architecture (number of layers, layer width, activation function)
beforehand.  
GPR also naturally includes
uncertainty quantification, which is an absolute necessity in 
experimental sciences. However, traditional GPR has mostly 
been derived and applied under an assumption of independent and identically distributed noise (i.i.d.~noise) \cite{hanuka2019online,williams2006gaussian,huang2006technical,schulz2017tutorial,mchutchon2011gaussian, stegle2011efficient,ballabio2019mapping}, i.e., noise 
that follows the same probability density function at each measurement point.
Since we are exclusively dealing with Gaussian statistics, this means that
all measurements have the same variance. 
In Kriging,
the geo-statistical analogue of GPR, this concept is called the nugget effect,
named after gold nuggets in the sub-surface. In early 
geo-statistical computations, the gold nuggets
lead to seemingly random errors. 
These were assumed to be constant across the domain. 
However, for materials-discovery experiments
the assumption of i.i.d.~noise is an unacceptable simplification. 
The variance of real experimental measurements vary greatly across the parameter space, and this has to be reflected in the steering 
process as well as in the final model creation. 
For instance, in x-ray scattering experiments, 
the variance of a raw measurement depends
strongly on the exposure time, computed quantities can have wildly different variances depending on the data in that part of the space (e.g. fit quality will not be uniform), and material heterogeneity will depend strongly on location within the parameter space.
These inhomogeneities in the measurement noise need to be actively included in the final
model to avoid interpolation mistakes and consequently erroneous models. 
Fortunately, non-i.i.d.~noise can easily be included in the GPR framework \cite{bijl2016gaussian,kuss2006gaussian}.
Large variances have to be countered with more measurements
in the respective areas until a desired uncertainty threshold is reached.
When creating the final model, 
the algorithm has to incorporate that the final model function
does not have to explain data points exactly if there is an associated variance.
Therefore, the model function does not have to pass through every data point.
After correct tuning, GPR is perfectly
equipped for this situation since it keeps track of a probability 
distribution over all possible model functions; conditioning will then produce
the most likely model function incorporating all measurement variances optimally. 

Another effect that has a 
significant impact on autonomous experiments is anisotropy of the parameter space, which is
either introduced by differing parameter 
ranges or different model variability in different parameter-space directions. 
In isotropic GPR one finds a single characteristic length scale for the data set. 
This was again motivated by early geo-statistical surveys 
in which isotropy was a good assumption.
However, when
one of the parameters is of significantly different magnitude, for instance
spatial directions in $\text{mm}~\in~[0,1]$ versus a temperature in $^\circ \mathrm{C}~\in~[5,500]$, we should find
different length scales for different directions of the parameter space. 
Also, there might be
different differentiability characteristics in different directions. 
It is therefore vitally important to give the model the flexibility 
to account for those varying features. This can either be done by 
using an altered Euclidean norm or by employing different norms that
provide more flexibility of distance measures in different directions.
The general idea including the concepts proposed in this paper are visualized in Figure
\ref{fig:schematic}.

This paper is organized as follows: First, we introduce the 
traditional theory of Gaussian process
regression with i.i.d.~noise and standard isotropic kernel functions. 
Second, we make formal changes to the theory to include non-i.i.d.
noise and anisotropy. 
Third, we demonstrate the impact of
the two concepts on synthetic experiments. 
Fourth, we present a synchrotron beamline experiment that
exploited both concepts in autonomous control.
\input{figures/figure1}
 
\section{Gaussian Process Regression with non-i.i.d.~Noise and Anisotropic Kernels}
\subsection{Prerequisite}
We define the parameter space $\mathcal{X} \subset \mathbb{R}^n$, 
which serves as the index set or input space in the scope of 
Gaussian process regression and elements
$\mathbf{x}~\in \mathcal{X}$. 
We define four functions over $\mathcal X$.
First, the latent function $f=f(\mathbf{x})$ can be interpreted as the
inaccessible ground truth.
Second, the often noisy measurements are described by 
$y=y(\mathbf{x}):~\mathcal{X}~\rightarrow~\mathbb{R}^d$. 
To simplify the derivation, 
we assume $d=1$; allowing for $d>1$ is a straightforward extension.
Third, the surrogate model function is then defined
as $\rho=\rho(\mathbf{x}):~\mathcal{X}~\rightarrow~\mathbb{R}$. 
Fourth, the posterior mean function $m(\mathbf{x})$, which is often assumed
to equal the surrogate model, i.e., $m(\mathbf{x})=\rho(\mathbf{x})$,
but this is not necessarily the case.
We also define a second space, a Hilbert space
$\mathcal{H}~\subset~\mathbb{R}^N~\times~\mathbb{R}^N~\times~\mathbb{R}^J$, with elements
$[\mathbf{f}~\mathbf{y}~\mathbf{f}_0]^T$,
where $N$ is the number of data points, $J$ is the number of points at 
which we want to predict the model function 
value, $\mathbf{y}$ are the measurement values, 
$\mathbf{f}$ is the vector of unknown latent function evaluations and 
$\mathbf{f}_0$ is the vector of predicted 
function values at a set of positions.
Note that scalar functions over $\mathcal{X}$, e.g. $f(\mathbf{x})$,
are vectors (bold typeface) in the Hilbert space $\mathcal{H}$, e.g. $\mathbf{f}$.
We also define a function $p$ over our Hilbert space which is just the
function value of the Gaussian probability density functions involved.
For more explanation on the distinction between the two spaces and the functions
involved see Figure \ref{fig:2spaces}.
\input{figures/figure2}
\subsection{Gaussian Process Regression with Isotropic Kernels and i.i.d.~Observation Noise}
Defining a GP regression model from data 
$D=\{(\mathbf{x}_1,y_1),...,(\mathbf{x}_N,y_N)\}$, 
where $y_i=f(\mathbf{x}_i)+\epsilon(\mathbf{x}_i)$,
is accomplished in a GP regression framework, by defining a 
Gaussian probability density function, called the prior, as
\begin{equation}
    p(\mathbf{f})=\frac{1}{\sqrt{(2\pi)^\mathrm{dim}|\mathbf{K}|}}
    \exp \left[ -\frac{1}{2}(\mathbf{f}-\pmb{\mu})^T \mathbf{K}^{-1}(\mathbf{f}-\pmb{\mu}) \right],
    \label{eq:prior}
\end{equation}
and a likelihood
\begin{equation}
    p(\mathbf{y})=\frac{1}{\sqrt{(2\pi)^\mathrm{dim}}\sigma}
    \exp \left[ -\frac{1}{2\sigma^2}(\mathbf{y}-\mathbf{f})^T (\mathbf{y}-\mathbf{f}) \right],
    \label{eq:likelihood_iid}
\end{equation}
where 
$\pmb{\mu}=[\mu(\mathbf{x}_1),...,\mu(\mathbf{x}_N)]^T$ is the mean of the prior
Gaussian probability density function (not to be confused with the posterior mean function $m(\mathbf{x})$). The prior mean can be understood as the position 
of the Gaussian.
$\mathbf{f}=[f(\mathbf{x}_1),...,f(\mathbf{x}_N)]^T$,  
$\mathbf{K}_{ij}~=~\mathbf{k}(\phi,\mathbf{x}_i,\mathbf{x}_j);~\mathbf{x}\in \mathcal{X}$ 
is the covariance of the 
Gaussian process, with its covariance function, often referred to as the kernel, 
$k(\phi,\mathbf{x}_i,\mathbf{x}_j)$, where $\phi$ are the hyper parameters, and 
where $\sigma^2$ is the variance of the i.i.d.~observation noise. The problem here is that,
in practice, the i.i.d.~noise restriction rarely holds in experimental sciences, 
which is one of the
issues to be addressed in this paper.
The kernel $k$ is a symmetric and positive semi-definite function, such that
$k:\mathcal{X} \times \mathcal{X}\rightarrow \mathbb{R}$.
As a reminder, $\mathcal{X}$ is our parameter space and often referred to as index set or input space in the literature. 
A well-known choice \cite{williams2006gaussian} is the Mat\'ern kernel class
defined by
\begin{align}
    k(\mathbf{x}_i,\mathbf{x}_j)~=~
    \sigma_s^2 \frac {2^{(1-\nu )}}{\Gamma (\nu )} \Bigg ( \sqrt {2\nu }\frac {r}{l }\Bigg )^{\nu }B_{\nu}\Bigg (\sqrt {2\nu }{\frac {r}{l}}\Bigg ),
    \label{eq:matern_kernel}
\end{align}
where $B_\nu$ is the Bessel function of second kind, $\Gamma$ is the gamma function,
$\sigma_s^2$ is the signal variance, $l$ is the length scale,
$r~=~||\mathbf{x}_i-\mathbf{x}_j||_{l_2}$ is the Euclidean distance between input points 
and $\nu$ is a parameter that controls the differentiability characteristics of
the kernel and therefore of the final model function.
The well-known exponential and squared exponential kernels are special cases
of the Mat\'ern kernels.
The signal variance $\sigma_s^2$ and the length scale $l$ are 
hyper parameters ($\phi$) that are found
by maximizing the log-likelihood, i.e., solving 
\begin{align}
        \argmax_{\phi~\mu}\Big(\log(L(\mathcal{D};\phi,\mu)) \Big)
    \label{eq:argmax_loglikelihood}                 
\end{align}
where
\begin{align}
    \log(L(\mathcal{D};\pmb{\phi},\mu(\mathbf{x})))~=\\ \nonumber
    &~-\frac{1}{2}(\mathbf{y}-\pmb{\mu}(\mathbf{x}))(\mathbf{K}(\phi)+\sigma^2~\mathbf{I})^{-1}(\mathbf{y}-\pmb{\mu}(\mathbf{x})) \\ \nonumber
    &~-\frac{1}{2}\log(|\mathbf{K}(\phi)+\sigma^2~\mathbf{I}|)
                         -\frac{\dim(\mathbf{y})}{2}\log(2\pi),
    \label{eq:loglikelihood}                 
\end{align}
where $I$ is the identity matrix.
In the isotropic case, we only have to optimize for one signal variance and one length scale (per kernel function).
The mean function $\mu(\mathbf{x})$ is often 
assumed to be constant and therefore does not have to be
part of the optimization. The mean function assigns the location of
the prior in $\mathcal{H}$ to any $\mathbf{x}~\in~\mathcal{X}$.
The mean function can therefore be used to communicate prior knowledge
(for instance physics knowledge) to the Gaussian process.
Provided some hyper parameters, the joint prior is given as
\begin{equation}
    p(\mathbf{f},\mathbf{f}_0)=\frac{1}{\sqrt{(2\pi)^\mathrm{dim}|\mathbf{\Sigma}|}}
    \exp \left[ -\frac{1}{2} \Big (
    \begin{bmatrix} \mathbf{f}-\pmb{\mu} \\  \mathbf{f}_0-\pmb{\mu}_0 \end{bmatrix}^T 
    \mathbf{\Sigma}^{-1}
    \begin{bmatrix} \mathbf{f}-\pmb{\mu} \\  \mathbf{f}_0-\pmb{\mu}_0 \end{bmatrix} \Big )\right ] ,
    \label{eq:joint_prior}
\end{equation}
where 
\begin{align}
   \Sigma~=~
   \begin{pmatrix}
   \mathbf{K}     & \pmb{\kappa} \\
   \pmb{\kappa}^T & \pmb{\mathcal{K}}
   \end{pmatrix},
   \label{eq:joint_covariance_matrix}
\end{align}
where $\pmb{\kappa}_i~=~k(\phi,\mathbf{x}_0,\mathbf{x}_i)$, 
$\pmb{\mathcal{K}}~=~k(\phi,\mathbf{x}_0,\mathbf{x}_0)$ and, as a reminder,
$\mathbf{K}_{ij}~=~k(\phi,\mathbf{x}_i,\mathbf{x}_j)$.
Intuitively speaking, $\Sigma$, $K$ and $k$ are all measures of similarity
between measurement results 
$y(\mathbf{x})$ of the input space. While $K$ stores this 
similarity between all data points, $\Sigma$ stores the similarity between all
data points and all unknown points of interest, and  $\kappa$ contains the 
similarity only between the unknown $y(\mathbf{x})$ of interest. 
$k$ contains the instruction on how to calculate this similarity.
The reader might wonder: ``how do we find the similarity between unknown points of
interest?'' The answer lies in the 
formulation of the kernels that calculate the
similarity just by knowing locations $\mathbf{x}~\in~\mathcal{X}$ and not the function 
evaluations $y(\mathbf{x})$.
$\mathbf{x}_0$ is the point where we want to estimate the mean and the variance.
Note here that, with only slight adaption of the equation, we are able to compute
the mean and variance for several points of interest.

The predictive distribution is defined as
\begin{align}\label{eq:pred_distr}
    p(\mathbf{f}_0|\mathbf{y})~&=~\int_{\mathbb{R^N}} 
    p(\mathbf{f}_0|\mathbf{f},\mathbf{y})~p(\mathbf{f},\mathbf{y})~d\mathbf{f} \nonumber \\
    &~\propto \mathcal{N}(\pmb{\mu}+\pmb{\kappa}^T~
    (\mathbf{K}+\sigma^2~\mathbf{I})^{-1}~(\mathbf{y}-\pmb{\mu}), \pmb{\mathcal{K}} -
    \pmb{\kappa}^T~(\mathbf{K}+\sigma^2~\mathbf{I})^{-1}~\pmb{\kappa})  
\end{align}
and the predictive mean and the predictive variance are therefore respectively
defined as
\begin{align}
    m(\mathbf{x}_0)&=\pmb{\mu} + \mathbf{k}^T(\mathbf{K}+\sigma^2~\mathbf{I})^{-1}(\mathbf{y}-\pmb{\mu}) \label{eq: pred_mean} \\ 
    \sigma^2(\mathbf{x}_0)&=k(\mathbf{x}_0,\mathbf{x}_0)- 
     \mathbf{k}^T(\mathbf{K} + \sigma^2~\mathbf{I})^{-1} \mathbf{k}, 
    \label{eq: pred_var}
\end{align}
which are the posterior mean and variance at $\mathbf{x}_0$, respectively. 
$\mathcal{N(\cdot,\cdot)}$ stands for the normal (Gaussian) distribution
with a given mean and covariance.
\subsection{Gaussian Processes with non-i.i.d.~Observation Noise}
To incorporate non-i.i.d.~observation noise one can redefine  
the likelihood \eqref{eq:likelihood_iid} as 
\begin{equation}
    p(\mathbf{y})=\frac{1}{\sqrt{(2\pi)^k|\mathbf{V}|}}
    \exp \left [ -\frac{1}{2}(\mathbf{y}-\mathbf{f})^T \mathbf{V}^{-1}(\mathbf{y}-\mathbf{f}) \right ] ,
    \label{eq:likelihood_noniid}
\end{equation}
where $\mathbf{V}$ is a diagonal matrix containing the
respective measurement variances. The matrix $\mathbf{V}$
can also have non-diagonal entries if the measurement noise happens to be correlated.
We will only discuss non-correlated measurement noise.

From equations \eqref{eq:joint_prior} and \eqref{eq:likelihood_noniid}, 
we can calculate
equation \eqref{eq:pred_distr}, i.e., 
the predictive probability distribution for a measurement 
outcome at $\mathbf{x}_0$, given the data set.
The mean and variance of this distribution are
\begin{align}
m(\mathbf{x}_0)&=\pmb{\mu} + \mathbf{k}^T(\mathbf{K}+\mathbf{V})^{-1}(\mathbf{y}-\pmb{\mu}) \label{eq:pred1} \\ 
\sigma^2(\mathbf{x}_0)&=k(\mathbf{x_0},\mathbf{x_0})- 
\mathbf{k}^T(\mathbf{K} + \mathbf{V})^{-1} \mathbf{k},\label{eq:pred2} 
\end{align}
respectively. 
Note here, that the matrix of the measurement errors $\mathbf{V}$ 
replaces the matrix $\sigma^2~\mathbf{I}$ in equations \eqref{eq: pred_mean}
and \eqref{eq: pred_var}.
However, this does not follow from a simple substitution, but from a
significantly different derivation. The log-likelihood \eqref{eq:loglikelihood}
changes accordingly, yielding
\begin{align}
    \log(L(\mathcal{D};\pmb{\phi},\mu(\mathbf{x})))~=\\ \nonumber
    &~-\frac{1}{2}(\mathbf{y}-\pmb{\mu})(\mathbf{K}(\phi)+\mathbf{V})^{-1}(\mathbf{y}-\pmb{\mu}) \\ \nonumber
    &~-\frac{1}{2}\log(|\mathbf{K}(\phi)+\mathbf{V}|)
     -\frac{\dim(\mathbf{y})}{2}\log(2\pi).
    \label{eq:loglikelihood}                 
\end{align}
This concludes the derivation of GPR with non-i.i.d.~observation noise.
Figure \ref{fig:non_iid} illustrates the effect of different kinds of noise on an one-dimensional model function.
As we can see, while some details of the derivation change when we account for
inhomogeneous (also known as input dependent or non-i.i.d) noise, the resulting
equation are very similar and the computation exhibits no extra costs.
\input{figures/figure3}
\subsection{Gaussian Processes with Anisotropy}
For parameter spaces $\mathcal{X}$ that are anisotropic, i.e., where different
directions have different characteristic correlation length, 
we can redefine the kernel function 
to incorporate different length scales in different directions.
One way of doing this for axial anisotropy 
is by choosing the $l^1$ norm as distance measure and redefine the kernel
function as
\begin{align}
    k(\mathbf{x}^m,\mathbf{x}^n) = \sigma_s^2~\prod_i^d~k_i(x^m_i~-~x^n_i;\phi_i),
\end{align}
where the superscripts $m,n$ mean point labels, the subscript $i$ means 
different directions in $\mathcal{X}$ and $d~=~\text{dim}(\mathcal{X})$.
Defining a kernel per direction gives us the flexibility to
enforce different orders of differentiability in different directions of $\mathcal{X}$.
The main benefit, however, is the possibility to define different length scales in
different directions of $\mathcal{X}$ (see Figure \ref{fig:anisotropic_kernels}). 
Unfortunately, the choice of the $l^1$ norm can lead to a very recognizable checkerboard 
pattern in the surrogate model, but the predictive power of the associated variance function is significantly improved compared to
the isotropic case. 

A second way, which avoids the checkerboard pattern in the model but does not allow different kernels in
different direction, is to redefine
the distances in $\mathcal{X}$ as 
\begin{align}
    r~=~\sqrt{\mathbf{x}^T~\mathbf{M}~\mathbf{x}},
\end{align}
where $\mathbf{M}$ is any symmetric positive semi-definite
matrix playing the role of a metric tensor \cite{vivarelli1999discovering}.
This is just the Euclidean distance in a transformed metric space.
In the actual kernel functions, any $r/l$ can then be replaced by the new equation for the metric.
We will here only consider axis-aligned anisotropy which means the matrix $\mathbf{M}$ is a diagonal matrix
with the inverse of the length scales on its diagonal. 
The extension to general forms of anisotropy is straightforward but needs a more costly likelihood optimization
since more hyper parameters have to be found. 
The rest of the theoretical treatment, however, remains
unchanged.
The mean function $\mu(\mathbf{x})$, the hyper parameters $\phi_i$ and the signal variance 
$\sigma_s^2$ are again found by 
maximizing the marginal log-likelihood \eqref{eq:loglikelihood}.
The associated optimization tries to find a maximum of a function 
that is defined over $\mathbb{R}^{d + 1}$, 
if we ignore the mean function as it is commonly done. We therefore 
have to find $d + 1$ parameters which adds a significant 
computational cost. 
If $\mathbf{M}$ is not diagonal we have to maximize 
the log-likelihood over $\mathbb{R}^{(d^2-N)/2 + 1}$.
However, the optimization can be performed in parallel 
to computing the posterior variance, 
which can hide the computational effort. 
It is important to note that accounting for anisotropy can make the training of the algorithm, i.e. the optimization of the log-likelihood, significantly more
costly. The extent of this depends on the kind of anisotropy considered.
As we shall see, taking anisotropy 
into account leads to more efficient steering and a higher-quality final result, and is thus generally worth the additional computational cost.
\input{figures/figure4}
\section{Synthetic Tests}
Our synthetic tests are carefully chosen to demonstrate the benefits
of the two concepts under discussion, namely: 
non-i.i.d.~observation noise and anisotropic kernels.
To demonstrate the importance of including non-i.i.d.~observation noise
into the analysis, we consider a synthetic test based 
on actual physics which we used in previous work to
showcase the functionality of past algorithms \cite{noack2019kriging}.
We are choosing an example given in a closed form, because it provides a
noise-free ``ground truth'' that we can compare to, whereas experimental data would inevitably include unknown errors. 
To showcase the importance of anisotropic kernels
as part of the analysis, we provide an high-dimensional example based on a simulation of a material that is subject to
a varying thermal history. 

The shown synthetic tests explore spaces of very different dimensionality.
There is no theoretical limitation to the
dimensionality of the parameter space.
Indeed the autonomous methods described herein are most advantageous 
when operating in high-dimensional spaces, since this is where simpler 
methods---and human intuition---typically fail to yield meaningful searches.
\subsection{Non-i.i.d.~Observation Noise}
For this test, we define a physical ``ground truth'' model ($f(\mathbf{x})$), 
whose correct function value at $\mathbf{x}$ is inaccessible due to 
non-i.i.d~measurement noise, but can be probed by our simulated experiment through $y(\mathbf{x})$. 
In this case, we assume that the measurements are subject to Gaussian noise 
with a standard deviation of $2\%$ of the function value at $\mathbf{x}$.
The ground-truth model function is defined to be the diffusion coefficient $D = D(r,T,C_m)$ for the 
Brownian motion of nanoparticles in a 
viscous liquid consisting of a binary mixture of water and glycerol: 
\begin{equation}
    D = \frac{k_B~T}{6\pi \mu r}, 
\label{eq:phys_model}
\end{equation}
where $k_B$ is Bolzmann's constant, $r~\in~[1~,100]~\mathrm{nm}$ is the nanoparticle radius, $T~\in~[0~,100]~^{\circ}\mathrm{C}$ is the temperature
and $\mu=\mu(T,C_m)$ is the viscosity as given by \cite{cheng2008formula}, where $C_m~\in~[0.0,100.0]~\%$ is the glycerol mass fraction. 
This model was used in \cite{noack2019kriging} to show the functionality of Kriging based autonomous experiments.
The experiment device has no direct access to the ground truth model, but adds an  unavoidable noise level, i.e.,
\begin{equation}
    D = \frac{k_B~T}{6\pi \mu r}~+~\epsilon(T,C_{m},r), 
\label{eq:phys_model1}
\end{equation}

To demonstrate the importance of the noise model, we first ignore noise $\epsilon$, then approximate it assuming i.i.d.~noise, and finally model it allowing for non-i.i.d.~noise.
Figure \ref{fig:3dnoise} shows the results after 500 measurements, and a comparison to the (inaccessible) ground truth.
Figure \ref{fig:3derror} compares the decrease in the error, in form of the Euclidean distance between the models and the ground truth, with increasing number of measurements $N$, between the three different types of noise.

The results show that treating noise as i.i.d.~or even non-existent 
can lead to artifacts in the surrogate model. Additionally, the discrepancy
between the ground truth and the surrogate mode is reduced far more efficiently if non-i.i.d.~noise is accounted for.
\vspace{1cm}
\input{figures/figure5}
\input{figures/figure6}
\subsection{Anisotropy}
Allowing anisotropy can increase efficiency of autonomous experiments significantly for
any dimensionality of the underlying parameter space.
However, as the dimensionality of the parameter space increases, 
the importance of  anisotropy increases substantially, purely due to the number of directions in which
anisotropy can occur.
To demonstrate this link, we simulated an experiment where a material is subjected to a varying thermal history. That is, the experiment consists of repeatedly changing the temperature, and taking measurements along this time-series of different temperatures. The temperature at each time step
can be thought of as one of the dimensions of the parameter space. The full set of possible applied thermal histories thus become points in the high-dimensional parameter space of temperatures. 

In particular, we consider the ordering of a block copolymer, which is a self-assembling material that spontaneously organizes into a well-defined morphology when thermally annealed \cite{doerk2017beyond}. The material organizes into a defined unit cell locally, with ordered grains subsequently growing in size as defects annihilate \cite{majewski2016rapid}. 
We use a simple model to describe this grain coarsening process, where the grain size $\xi$ increases with time according to a power-law
\begin{equation}
    \xi = k t^{\alpha} ,
\label{eq:bcp_grain}
\end{equation}
where $\alpha$ is a scaling exponent (set to $0.2$ for our simulations) and the prefactor $k$ captures the temperature-dependent kinetics
\begin{equation}
    k = A e^{-E_a/k_B T} .
\label{eq:bcp_Arrhenius}
\end{equation}
Here, $E_a$ is an activation energy for coarsening (we select a typical value of $E_a = 100 \, \mathrm{kJ/mol}$), and the prefactor $A$ sets the overall scale of the kinetics (set to $3 \times 10^{11} \, \mathrm{nm}/\mathrm{s}^{\alpha}$). 
From these equations we construct an instantaneous growth-rate of the form:
\begin{equation}
    \frac{\mathrm{d}\xi}{\mathrm{d}t} = k^{1/\alpha} \xi^{1-1/\alpha} .
\label{eq:bcp_growth}
\end{equation}
Block copolymers are known to have a order-disorder transition temperature ($T_{\mathrm{ODT}}$) above which thermal energy overcomes the material's segregation strength, and thus the nanoscale morphology disappears in favor of a homogeneous disordered phase. Heating beyond $T_{\mathrm{ODT}}$ thus implies driving $\xi$ to zero. We describe this `grain dissolution' process using an ad-hoc form of:
\begin{equation}
    \frac{\mathrm{d}\xi}{\mathrm{d}t} = - k_{\mathrm{diss}} (T-T_{\mathrm{ODT}}) ,
\label{eq:bcp_dissolve}
\end{equation}
where we set $k_{\mathrm{diss}} = 1.0 \, \mathrm{nm\,s^{-1}\,K^{-1}}$ and $T_{\mathrm{ODT}} = 350 \, ^{\circ}\mathrm{C}$. We also apply ad-hoc suppresion of kinetics near $T_{\mathrm{ODT}}$ and when grain sizes are very large to account for experimentally-observed effects. 
Overall, this simple model describes a system wherein grains coarsen with time and temperature, but shrink in size if temperature is raised too high. The parameter space defined by a sequence of temperatures will thus exhibit regions of high or low grain size depending on the thermal history described by that point; moreover there is non-trivial coupling between these parameters since the grain size obtained for a given step of the annealing (i.e. a given direction in the parameter space) sets the starting-point for coarsening in the next step (i.e. the next direction of the parameter space).

We select thermal histories consisting of 11 temperature selections (temperature is updated every $6 \, \mathrm{s}$), which thus defines an 11-dimensional parameter space for exploration. 
Each temperature history defines a point ($\mathbf{x}~\in~\mathcal{X}$) within the 11-dimensional input space.
As can be seen in Figure \ref{fig:11d}(a), the majority of thermal histories one might select terminate in a relatively small grain size (blue lines in figure). 
This can be easily understood since a randomly-selected annealing protocol will use temperatures that are either too low (slow coarsening) or too high ($T>T_{\mathrm{ODT}}$ drives into disordered state). Only a subset of possible histories terminate with a large grain size (dark, less transparent lines in Figure \ref{fig:11d}), corresponding to the judicious choice of annealing history that uses large temperatures without crossing ODT. 
While this conclusion is obvious in retrospect, in the exploration of a new material system (e.g. for which the value of material properties like $T_{\mathrm{ODT}}$ are not known), identifying such trends is non-trivial. Representative slices through the 11-dimensional parameter space (Fig. \ref{fig:11d}(b) and (c)) further emphasize the complexity of the search problem, especially emphasizing the anisotropy of the problem. That is, different steps in the annealing protocol have different effects on coarsening; correspondingly the different directions in the parameter space have different characteristic length scales that must be correctly modeled (even though every direction is conceptually similar in that it describes a $6 \, \mathrm{s}$ thermal annealing process).

Autonomous exploration of this parameter space enables the construction of a model for this coarsening process. Moreover, the inclusion of anisotropy markedly improves the search efficiency, reducing the model error more rapidly than when using a simpler isotropic kernel (Fig. \ref{fig:11d}(d)). As the dimensionality of the problem and the complexity of the physical model increase, the utility of including an anisotropic kernel increases further still.
\input{figures/figure7}
\section{Autonomous SAXS Exploration of Nanoscale Ordering in a Flow-Coated Polymer-Grafted Nanorod Film}
The proposed GP-driven decision-making algorithm that takes into account non-i.i.d.~observation
noise and anisotropy has been used successfully in autonomous synchrotron experiments. 
Here we present, as an illustrative example, the results of an autonomous x-ray scattering experiment on a polymer-grafted gold nanorod thin film, where a combinatorial sample library was used to explore the effects of film fabrication parameters on self-assembled nanoscale structure.

Unlike traditional short ligand coated particles, polymer-grafted nanoparticles (PGNs) are stabilized by high molecular weight polymers at relatively low grafting densities. As a result, PGNs behave as soft colloids, possessing the favorable processing behavior of polymer systems while still retaining the ability to pack into ordered assemblies \cite{che2016preparation}. Although this makes PGNs well suited to traditional approaches for thin-film fabrication, the nanoscale assembly of these materials is inherently complex, depending on a number of variables including, but not limited to particle-particle interactions, particle-substrate interactions, and process methodology. 

The combinatorial PGN film sample was fabricated at the Air Force Research Laboratory. A flow-coating method \cite{che2016preparation} was used to deposit a thin PGN film on a surface-treated substrate where gradients in coating velocity and substrate surface energy were imposed along two orthogonal directions over the film surface. A 250 nM toluene solution of 53 kDa polystyrene-grafted gold nanorods (94\% polystyrene by volume), with nanorod dimensions of $70 \pm 6$ nm in length and $11.0 \pm 0.9$ nm in diameter (based on TEM analysis), was cast onto a functionalized glass coverslip using a motorized coating blade. The resulting film covered a rectangular area of dimensions 50 mm $\times$ 60 mm. The surface energy gradient on the glass coverslip was generated through the vapor deposition of phenylsilane \cite{genzer2002molecular}. The substrate surface energy varied linearly along the $x$ direction from 30.5 mN/m (hydrophobic) at one edge of the film ($x=0$) to 70.2 mN/m (hydrophilic) at the other edge ($x=50$ mm). Along the $y$ direction, the film-casting speed increased from 0 mm/s (at $y=0$) to 0.5 mm/s ($y=60$ mm) at a constant acceleration of 0.002 mm/s$^{2}$. The film-casting condition corresponds to the evaporative regime where solvent evaporation occurs at similar timescales to that of solid film formation \cite{bao2018meniscus}. In this regime, solvent evaporation at the meniscus induces a convective flow, driving the PGNs to concentrate and assemble at the contact line. The film thickness decreased with increasing coating speed, resulting in transitions from multilayers through a monolayer to a sub-monolayer with increasing $y$. This was verified by optical microscopy observations of the boundaries between multilayer, bilayer, monolayer and sub-monolayer regions, the last of which were identified by the presence of holes in the film, typically 1 $\mu$m or greater as seen in the optical images.

The autonomous small-angle x-ray scattering (SAXS) experiment was performed at the Complex Materials Scattering (11-BM CMS) beamline at the National Synchrotron Light Source II (NSLS-II), Brookhaven National Laboratory. As described previously \cite{noack2019kriging, noack2020advances}, experimental control was coordinated by combining three Python software processes: \textit{bluesky} \cite{bluesky} for automated sample translations and data collection,  \textit{SciAnalysis} \cite{SciAnalysis} for real-time analysis of newly collected SAXS images, and the above GPR-based optimization algorithms for decision-making. The incident x-ray beam was set to a wavelength of 0.918 \AA\, (13.5 keV x-ray energy) and a size of 0.2 mm $\times$ 0.2 mm. The PGN film-coated substrate was mounted normal to the incident x-ray beam, on a set of motorized $xy$ translation stages. Transmission SAXS patterns were collected on an area detector (DECTRIS Pilatus 2M) located at a distance of 5.1 m downstream of the sample, with exposure time of 10 s/image. The SAXS results indicate that the polymer grafted nanorods tend to form ordered domains in which the nanorods lie flat and parallel to the surface and align with their neighbors. The fitting of SAXS intensity profiles via real-time analysis allowed for the extraction of quantities such as: the scattering-vector position $q$ for the diffraction peak corresponding to the in-plane inter-nanorod spacing $d = 2\pi/q$; the degree of anisotropy $\eta \in [0, 1]$ for the in-plane inter-nanorod alignment, where $\eta = 0$ for random orientations and $\eta = 1$ for perfect alignments \cite{ruland2004saxs}; the azimuthal angle $\chi$ or the factor $\cos(2\chi)$ for the in-plane orientation of the inter-nanorod alignment; and the grain size $\xi$ of the nanoscale ordered domains, which is inversely proportional to the diffraction peak width and provides a measure of the extent of in-plane positional correlations between aligned nanorods. The analysis-derived best-fit values and associated variances for these parameters were passed to the GPR decision algorithms.

Three analysis-derived quantities $\xi$, $\eta$, and $\cos(2\chi)$ were used as signals to steer the SAXS measurements as a function of surface coordinates $(x, y)$. For the initial part of the experiment, $N < 464$ (first 4 h), where $N$ is the number of measurements completed up to a given point in the experiment, the autonomous steering utilized the exploration mode based on model uncertainty maxima \cite{noack2020advances} for $\xi$, $\eta$, and $\cos(2\chi)$. For the latter part of the experiment ($464 \leq N \leq 1520$ or next 11 h), the feature maximization mode \cite{noack2020advances} was used for $\eta$, while keeping $\xi$ and $\cos(2\chi)$ in the exploration mode. We found that the nanorods in the ordered domains tended to orient such that their long axes were aligned along the $x$ direction [$\cos(2\chi) \approx 1$], i.e.,  perpendicular to the coating direction, and that $\xi$ and $\eta$ are strongly coupled. Figure \ref{fig:time_evo}A (top panels) show the $N$-dependent evolution of the model for the grain size distribution $\xi$ over the film surface. It should be noted that the entire experiment took 15 h, and that the GPR-based autonomous algorithms identified the highly ordered regions in the band $5 < y < 15\, mm$ (between red lines in Fig. \ref{fig:time_evo}A), corresponding to the uniform monolayer region, within the first few hours. By contrast, grid-based scanning-probe transmission SAXS measurements would not be able to identify large regions of interest at these resolutions in such a short amount of time.

The collected data is corrupted by non-i.i.d.~measurement noise.
While all signals are corrupted by noise, we draw attention to the peak position $q$ because it shows the most
obvious correlation of non-i.i.d.~measurement noise and model certainty. The green circles in Figure \ref{fig:time_evo}B (middle panel) and C (right panel) highlight
the areas where the measurement noise affects the Gaussian-process predictive variance significantly. Note that we have not used $q$ for steering in this case, but the general principle we want to show remains unchanged across all experiment results. 
Figure \ref{fig:time_evo}A shows the 
time evolution of the exploration of the model and the impact of non-i.i.d.~noise on the model but also
on the uncertainty. If $q$ had been used for steering without taking into account non-i.i.d.noise into the analysis, 
the autonomous experiment would have been misled because predictive uncertainty due to high noise levels would not have been
taken into account. Figure \ref{fig:time_evo} shows 
that the next suggested measurement strongly depends on the noise. We want to remind the reader at this point that the next optimal measurement happens at the maximum of the GP predictive variance. The locations of the optima (Figure \ref{fig:time_evo}C) are clearly different when non-i.i.d.~noise is taken into account. 
The objective function without measurement noise
(Fig. \ref{fig:time_evo}C, left panel) shows no preference for regions of high noise (green circles in Fig. \ref{fig:time_evo}B, middle panel), where preference means higher function values of the GP predictive variance. In contrast, the variance function that takes measurement 
noise into account (Fig. \ref{fig:time_evo}C, right panel) gives preference to regions (green circles) 
where measurement noise of the data is high. This is a significant advantage and can only be
accomplished by taking into account non-i.i.d.~measurement noise. In 
conclusion, the model that assumes no noise looks better resolved, which communicates a wrong level 
of confidence and misguides the steering. The model that takes into account non-i.i.d.~noise 
finds the correct most likely model and the corresponding uncertainty. 
The algorithm also took advantage of anisotropy by learning a slightly longer 
length scale in the x-direction which increased the overall model certainty.
Note that the algorithm used an objective function formulation that put emphasis on high-amplitude
regions of the parameter space. This led to a higher resolution in areas of interest.

The above autonomous SAXS experiment revealed interesting features from the material fabrication perspective as well. First, a somewhat surprising result is that the grain size is not observed to change significantly with surface energy (Figure \ref{fig:time_evo}A). Previous work on the assembly of polystyrene-grafted spherical gold nanoparticles \cite{che2016preparation} demonstrated a significant decrease in nanoparticle ordering when fabricating films on lower surface energy substrates (greater polymer-substrate interactions). Although the surface energies used in this study are similar, a different silane was used to modify the glass surface (phenylsilane vs octyltrichlorosilane) which may differ in its interaction with polystyrene. We also note that PGN-substrate interactions will be sensitive to molecular orientation of the functional groups, which is known to be highly dependent on the functionalization procedure \cite{genzer2002molecular}. Second, an unexpected well-ordered band was identified at $20 < x < 35$ mm and $y > 15$ mm (between blue lines in Figure \ref{fig:time_evo}A), corresponding to the sub-monolayer region with an intermediate surface-energy range. We believe that this effect arises from instabilities associated with the solution meniscus near the middle of the coating blade ($x \sim 25$ mm). Rapid solvent evaporation often leads to undesirable effects including the generation of surface tension gradients, Marangoni flows, and subsequent contact line instabilities. This can result in the formation of non-uniform morphologies as demonstrated by the irregular region of larger grain size centered in the middle of the film and spanning the entire velocity range. Further investigations into these issues are currently in progress.

\input{figures/figure8}
\section{Discussion and Conclusion}
In this paper, we have demonstrated the importance of including inhomogeneous (i.e. non-i.i.d.)
observation noise and anisotropy into Gaussian-process-driven autonomous materials-discovery experiments.

It is very common in the scientific community to rely on Gaussian processes that ignore measurement noise
or only include homogeneous noise, i.e. noise that is a constant for every measurement. In experimental sciences,
and especially in experimental material sciences, strong inhomogeneity in measurement noise can be present and 
only accounting for homogeneous (i.i.d) measurement noise is therefore 
insufficient and leads to inaccurate models and,
in the worst case, wrong interpretations and missed scientific discoveries. 
We have shown that it is straightforward to include non-i.i.d noise into the steering and modeling
process. Figure \ref{fig:3dnoise} undoubtedly shows the benefit of including non-i.i.d measurement
noise into the Gaussian process analysis. Figure \ref{fig:3derror} supports the conclusion
we drew from Figure \ref{fig:3dnoise} visually, by showing a faster error decline.

The case for allowing anisotropy in the input space can be made when there is a reason to believe
that data varies much more strongly in certain direction than in others. This is often the case when
the directions have different fundamentally physical meanings. For instance, one direction can
mean a temperature, while another one can define a physical distance. In this case, accounting for anisotropy
can be vastly beneficial, since the Gaussian process will learn the different length scales and use them
to lower the overall uncertainty. Figure \ref{fig:11d} shows how common anisotropy is, even in cases
where it would normally not be expected, and how including it decreases the approximated 
error of the Gaussian process posterior
mean. In our example, all axes carry the unit of temperature; even so, anisotropy is present and accounting for
it has a significant impact on the approximation error. 

In our autonomous synchrotron x-ray experiment we have seen how misleading the no-measurement-noise can be.
While the Gaussian process posterior mean, assuming no noise, is much more detailed in Figure \ref{fig:time_evo},
it is not supported by the data which is subject to non-i.i.d.~noise. In addition, we have seen that
the steering actually accounts for the measurement noise if included, which leads to much a smarter decision algorithm 
that knows where data is of poor quality and has to be substantiated. We showed, that without accounting for non-i.i.d.~noise
this phenomenon would not arise. We would therefore place measurements sub-optimally, wasting
device access, staff time and other resources.

It is important to discuss the computational costs that come with accounting for non-i.i.d.~noise and
anisotropy. While non-i.i.d.~noise can be included at no additional computational costs, anisotropy potentially
comes at a price. The more complex the anisotropy, the more hyper parameters have to be found. The number of hyper parameters
translates directly into the dimensionality of the space the likelihood is defined over. The training process
to find the hyper parameters will therefore take longer, the more hyper parameters we have to find. However, the cost
per function evaluation will not change significantly. Therefore, instead of avoiding the valuable anisotropy, we should
make use of modern, efficient optimization methods. 

While our results have shown that accounting for non-i.i.d.~noise and anisotropy is highly valuable
for the efficiency of an autonomously steered experiment, we have only scratched the surface of possibilities.
Both proposed improvements can be seen as part of a larger theme commonly referred to as kernel design.
The possibilities for improvements and tailoring of Gaussian-process-driven steering of experiments
are vast. Well-designed kernels have the power to extract sub-spaces of the Hilbert space of functions,
which means we can put constraints on the function we want to consider as our model.
We will look into the impact of advanced kernel designs 
on autonomous data acquisition in the near future.

\section{Acknowledgments}
The work was partially funded through the Center for Advanced
Mathematics for Energy Research Applications 
(CAMERA), which is jointly funded by the 
Advanced Scientific Computing Research (ASCR) 
and Basic Energy Sciences (BES) within the 
Department of Energy's Office of Science, under Contract No. DE-AC02-05CH11231. 
This work was conducted at Lawrence Berkeley National Laboratory and 
Brookhaven National
Laboratory. This research used resources of the Center for Functional
Nanomaterials and the National Synchrotron 
Light Source II, which are U.S. DOE
Office of Science Facilities, at Brookhaven National 
Laboratory under Contract
No. DE-SC0012704.
Partial funding was supplied by the Air Force Research Laboratory Materials and Manufacturing Directorate and the Air Force Office of Scientific Research.

\section*{Author Contributions Statement}
M.N., K.G.Y., and 
M.F. developed the key ideas. M.N. devised the necessary algorithm, 
formulated the required mathematics, and implemented the computer codes. 
M.F. and K.G.Y. 
designed the x-ray scattering experiment. R.A.V. conceived the material and process design. J.K.S. prepared 
the samples and performed preliminary characterizations. M.N., K.G.Y., M.F., G.D., and R.L. performed the autonomous experiments. K.G.Y. analyzed the experimental data. M.N. analyzed the 
algorithm performance and wrote the first draft of the manuscript. M.F. and K.G.Y. supervised the work. All authors discussed the results 
and commented on the manuscript.
\section*{Additional Information}
\subsection*{Competing Interests}
The authors declare no competing interests.
\clearpage
\bibliographystyle{plainnat}
\bibliography{./literature}
\end{document}

%% file: figures/figure1.tex
\begin{figure}[htbp]
  \centering
   \input{./figures/GeneralSetup.pdf_tex}
\begin{subfigure}[t]{0.32\linewidth}
\includegraphics[trim={3cm 2cm 3cm 1cm},clip,width = \linewidth]{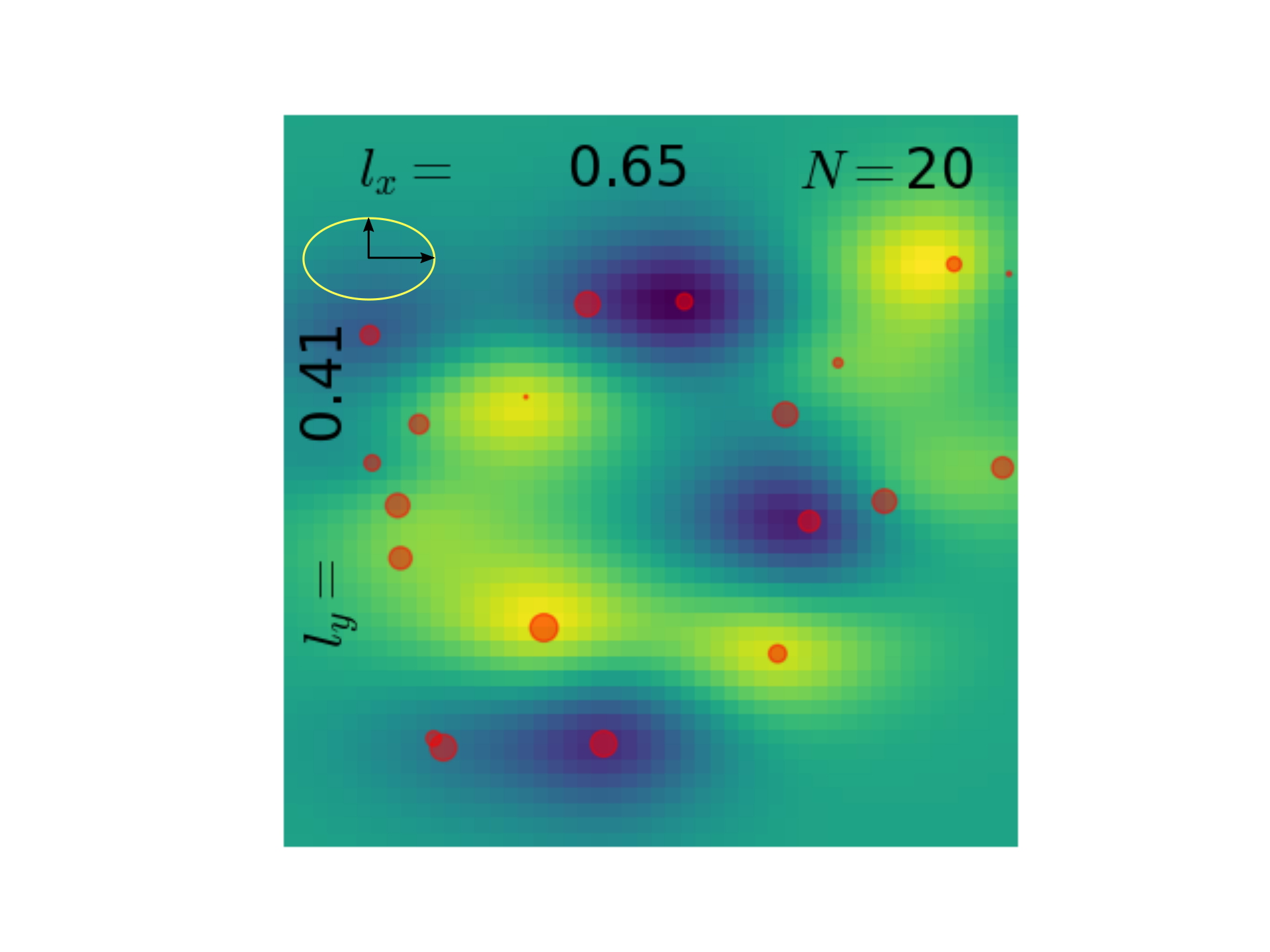}
\end{subfigure}
 \begin{subfigure}[t]{0.32\linewidth}
\includegraphics[trim={3cm 2cm 3cm 1cm},clip,width = \linewidth]{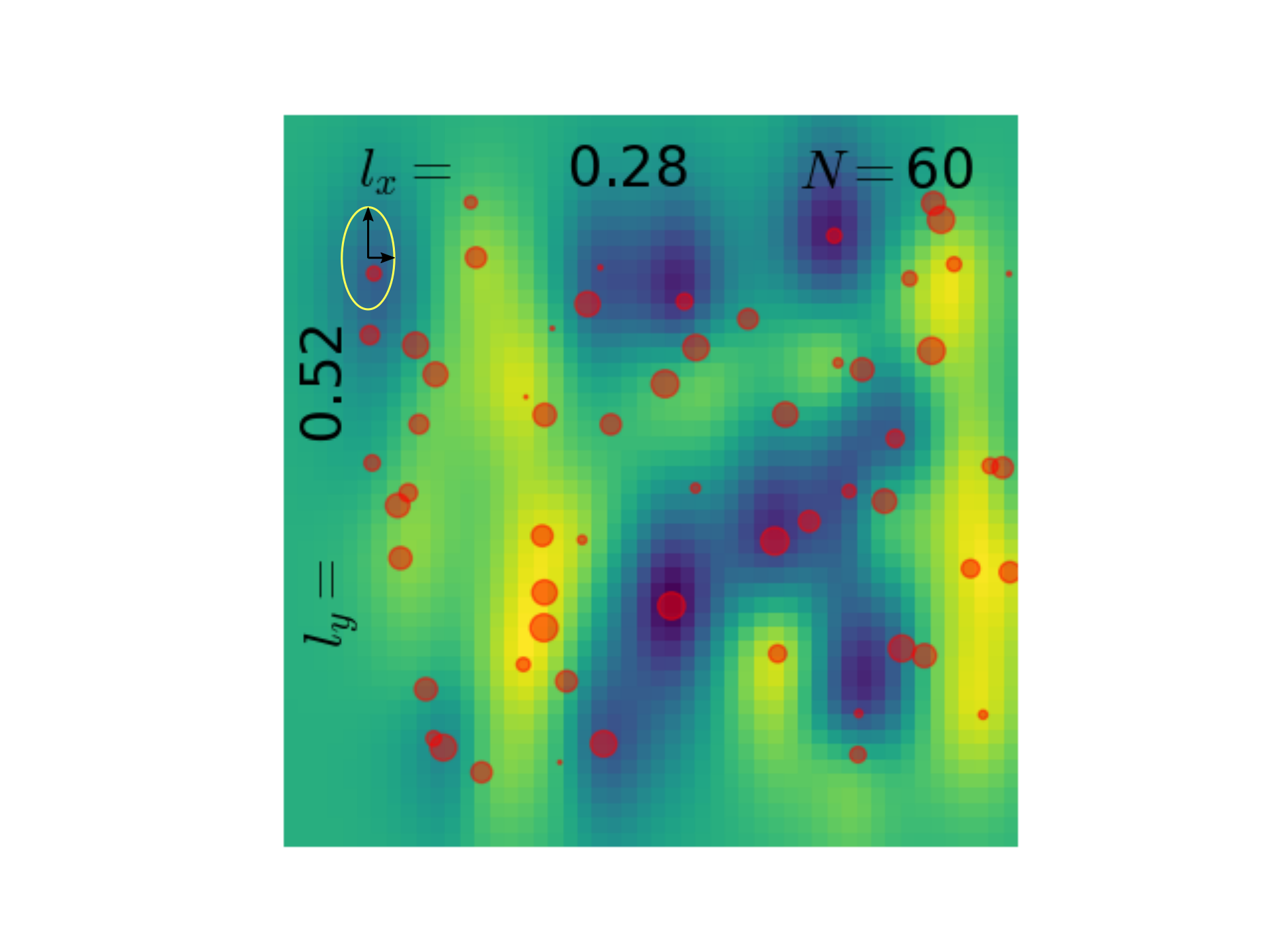}
\end{subfigure}
 \begin{subfigure}[t]{0.32\linewidth}
\includegraphics[trim={3cm 2cm 3cm 1cm},clip,width = \linewidth]{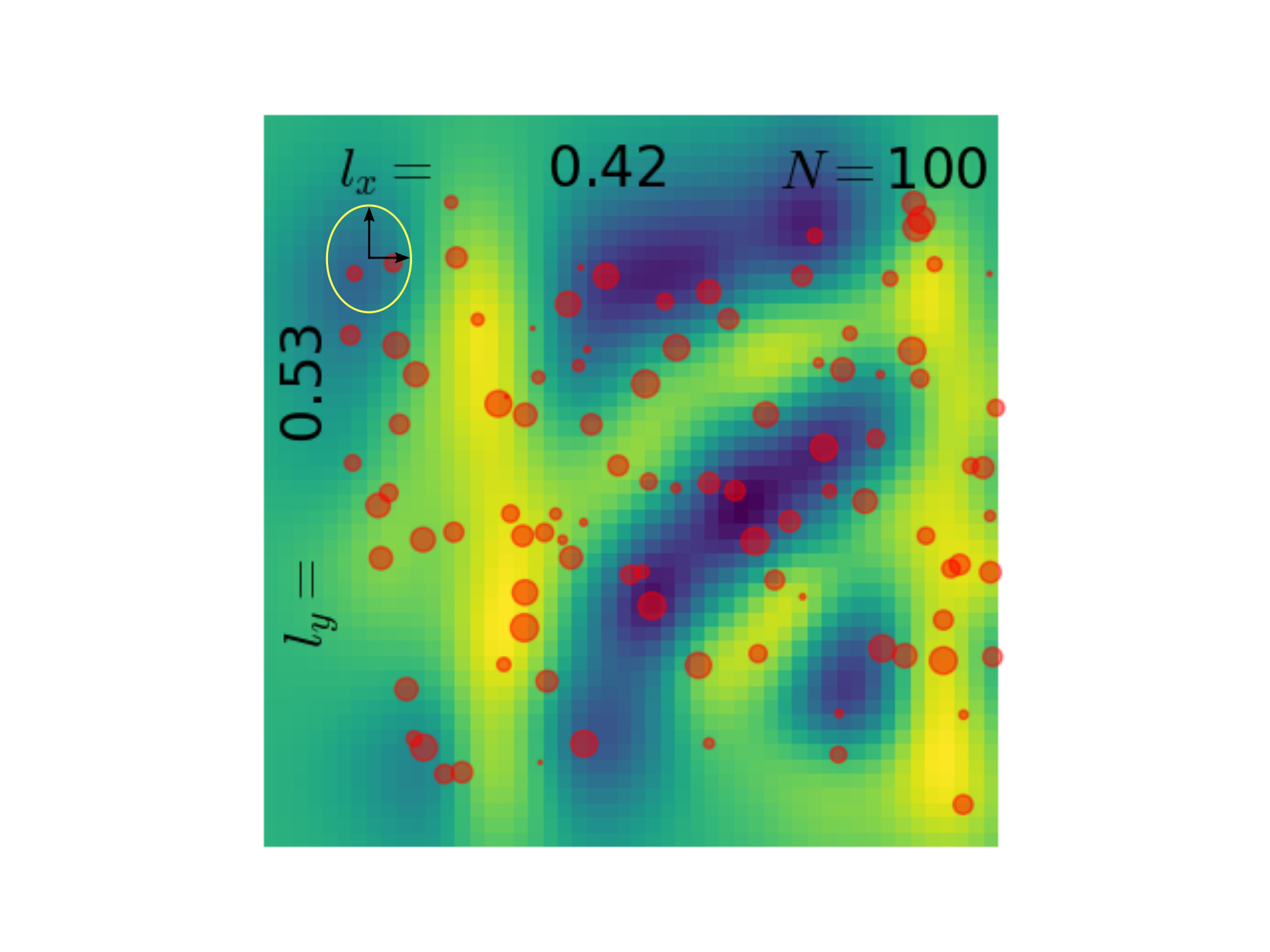}
\end{subfigure}
\caption{
Schematic of an autonomous experiment.
The data acquisition device in this example is a beamline at a synchrotron light source.
The measurement result depends on parameters $\mathbf{x}$.
The raw data is then sent through an automated data processing and analysis pipeline.
From the analyzed data, 
the autonomous-experiment algorithm creates a surrogate model and an uncertainty
function whose maxima represent points of high-value measurements; they are found by employing
function optimization tools. The new measurement parameters $\mathbf{x}$ are then 
communicated to the data acquisition device and the loop starts over.
The main contribution of the present work is that the model computation and uncertainty quantification account for the anisotropic nature of the 
model function and
the input-dependent (non-i.i.d.) measurement noise.
The surrogate model (bottom) shows how the model function is evolving
as the experiment is steered and more data ($N$) is collected. 
The red dots indicate the positions of the measurements and their size represents the
varying associated measurement variances.
The numbers $l_x$ and $l_y$ indicate the anisotropic correlation lengths
that the algorithm finds by maximizing a log-likelihood function. 
The ellipses show the found anisotropy visually. 
The take-home message for the practitioner here is that the method will 
find the most likely model function given all collected data with their variances. 
The model function will not pass directly through the points but find 
the most likely shape given all available information. 
}
\label{fig:schematic}
\end{figure}

%% file: figures/GeneralSetup.pdf_tex
\begingroup%
  \makeatletter%
  \providecommand\color[2][]{%
    \errmessage{(Inkscape) Color is used for the text in Inkscape, but the package 'color.sty' is not loaded}%
    \renewcommand\color[2][]{}%
  }%
  \providecommand\transparent[1]{%
    \errmessage{(Inkscape) Transparency is used (non-zero) for the text in Inkscape, but the package 'transparent.sty' is not loaded}%
    \renewcommand\transparent[1]{}%
  }%
  \providecommand\rotatebox[2]{#2}%
  \ifx\svgwidth\undefined%
    \setlength{\unitlength}{545.52729492bp}%
    \ifx\svgscale\undefined%
      \relax%
    \else%
      \setlength{\unitlength}{\unitlength * \real{\svgscale}}%
    \fi%
  \else%
    \setlength{\unitlength}{\svgwidth}%
  \fi%
  \global\let\svgwidth\undefined%
  \global\let\svgscale\undefined%
  \makeatother%
  \begin{picture}(1,0.45344618)%
    \put(0.45,0.3){x-ray beam}%
    \put(0.50,0.45){\large data acquisition device}%
    \put(0.50,0.42){\large e.g. a synchrotron light source}%
    \put(0.30,0.32){\large sample $f(\mathbf{x})$}%
    \put(0.30,0.30){\large surrogate $\rho(\mathbf{x})$}%
    \put(0.0,0.28){\large detector image}%
    \put(0.165,0.11){\large automated data}
    \put(0.185,0.09){\large analysis}
    \put(0.36,0.16){\large model}
    \put(0.33,0.14){\large computation}
    \put(0.335,0.04){\large uncertainty}
    \put(0.325,0.02){\large quantification}
    \put(0.65,0.125){\large objective}
    \put(0.65,0.105){\large function}
    \put(0.65,0.25){\large function}
    \put(0.63,0.23){\large optimization}
    \put(0.7,0.32){\large communicate}
    \put(0.7,0.3){\large measurement}
    \put(0.7,0.28){\large requests}
    \put(0,0){\includegraphics[width=\linewidth,page=1]{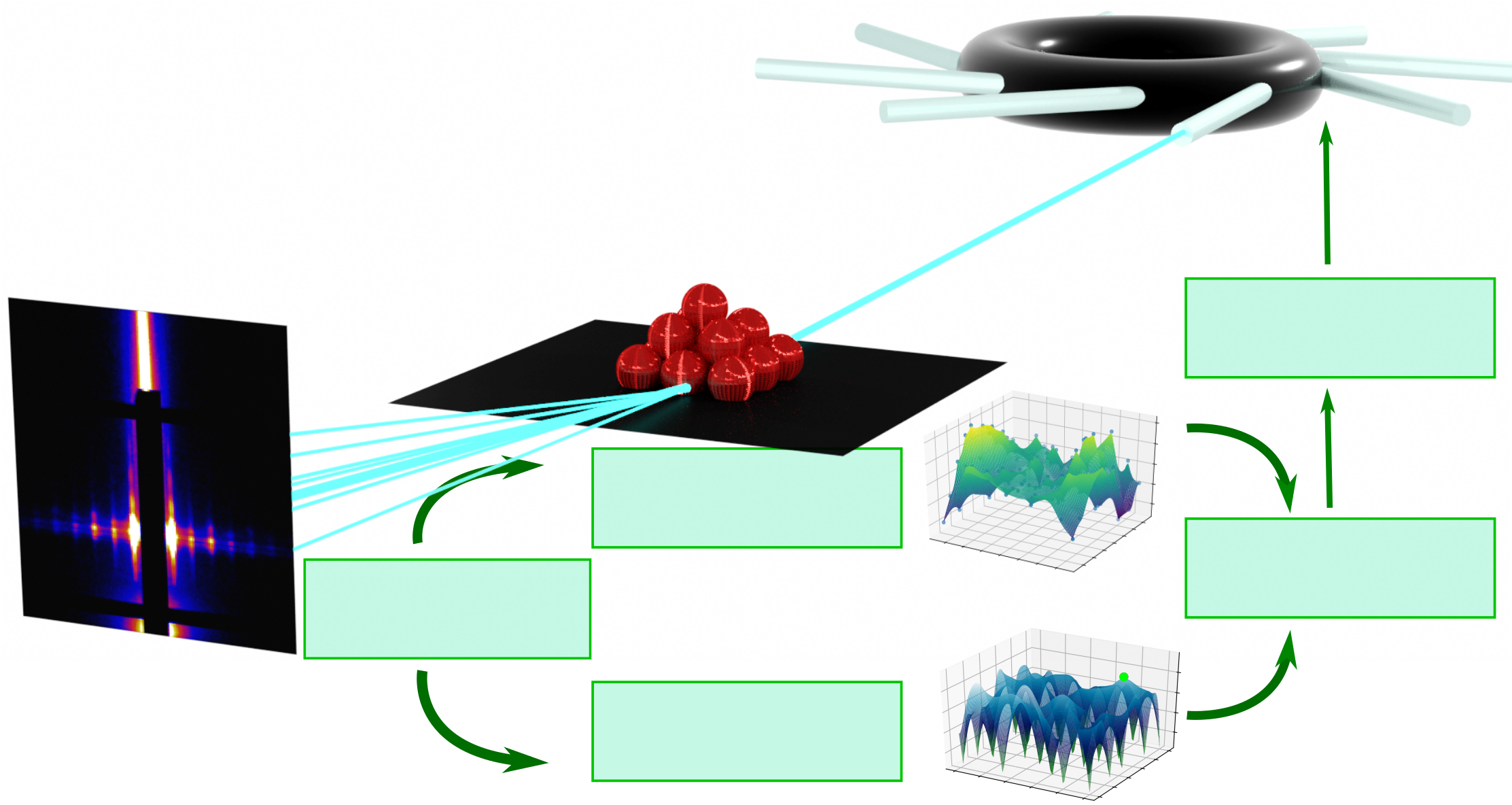}}%
  \end{picture}
\endgroup%

%% file: figures/figure2.tex
\begin{figure}[htbp]
    \centering
    \begin{subfigure}[t]{0.48\linewidth}
    \includegraphics[trim={2cm 1cm 1cm 1cm},clip,width = \linewidth]{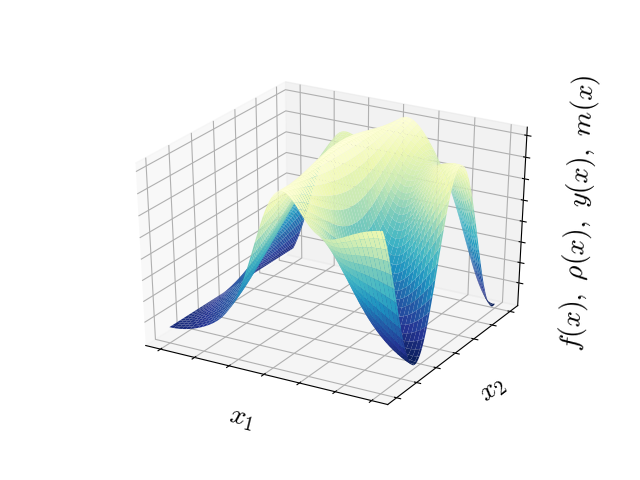}
    \caption{}
    \end{subfigure}
        \begin{subfigure}[t]{0.48\linewidth}
    \includegraphics[trim={2cm 1cm 1cm 1cm},clip,width = \linewidth]{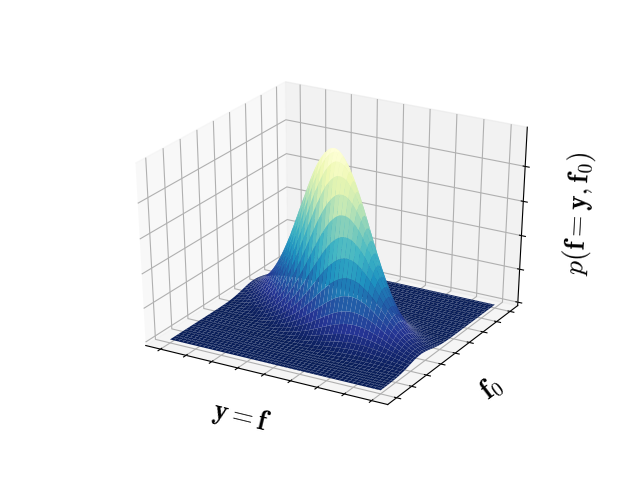}
    \caption{}
    \end{subfigure}
    \caption{Figure emphasizing the distinction between the spaces and functions involved in the derivation. (a) A function over $\mathcal{X}$. 
    This can be the surrogate model $\rho(\mathbf{x})$, 
    the latent function $f(\mathbf{x})$ to be approximated through an experiment, the function describing the measurements
    $y(\mathbf{x})$ or the predictive mean function $m(\mathbf{x})$. 
    $x_1$ and $x_2$ are two experimentally controlled parameters (e.g., synthesis, 
    processing or environmental conditions) that the measurement outcomes potentially depend on. (b) The Gaussian probability density function over $\mathcal{H}$ which gives
    GPR its name. For noise-free measurements, $\mathbf{y}~=~\mathbf{f}$
    at measurement points, meaning that we can directly observe the model function. Generally this is not the case and the observations
    $\mathbf{y}$ are corrupted by input dependent (non-i.i.d) noise.}
    \label{fig:2spaces}
\end{figure}

%% file: figures/figure3.tex
\begin{figure}[htbp]
    \centering
    \begin{subfigure}[t]{0.32\linewidth}
    \includegraphics[trim={1.0cm 0.5cm 1.5cm 0.5cm},clip,width = \linewidth]{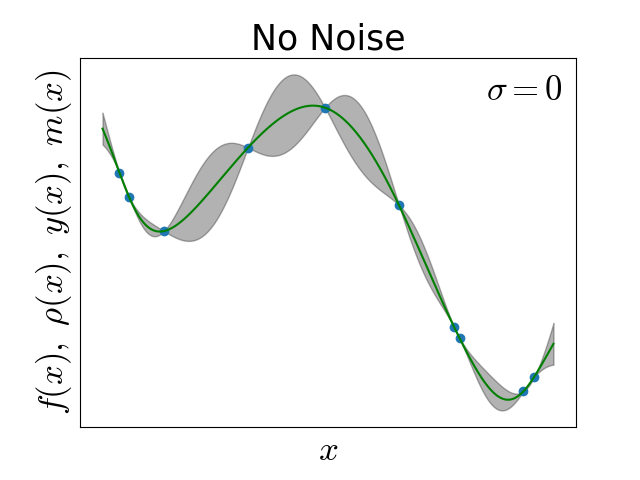}
    \caption{}
    \end{subfigure}
    \begin{subfigure}[t]{0.32\linewidth}
    \includegraphics[trim={1.0cm 0.5cm 1.5cm 0.5cm},clip,width = \linewidth]{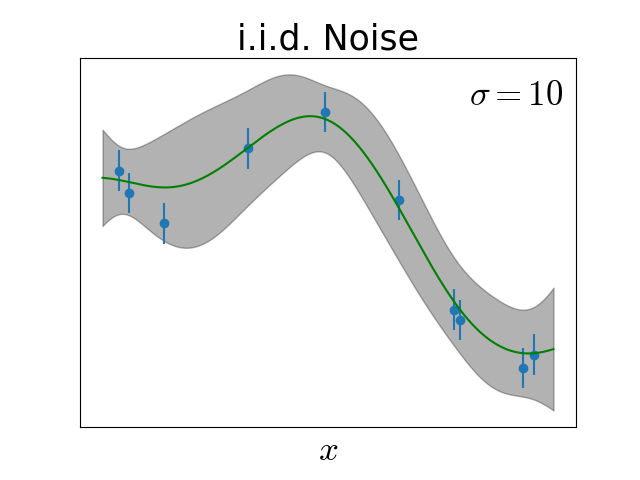}
    \caption{}
    \end{subfigure}
    \begin{subfigure}[t]{0.32\linewidth}
    \includegraphics[trim={1.0cm 0.5cm 1.5cm 0.5cm},clip,width = \linewidth]{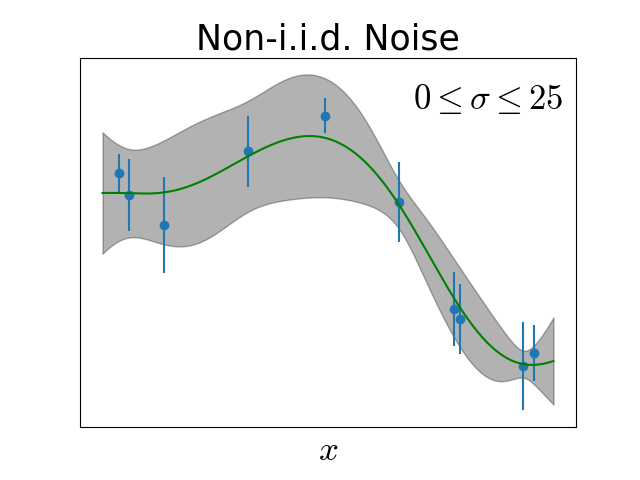}
    \caption{}
    \end{subfigure}
    \caption{Three one-dimensional examples with (a) no noise, 
    (b) i.i.d.~noise and (c) non-i.i.d.~noise, respectively. For the no-noise case, the model has to explain the data exactly.
    In the i.i.d.~noise-case, the algorithm is free to choose a model that does not explain the data exactly but allows 
    for a constant measurement variance. In the non-i.i.d.~noise case, the algorithm finds the most likely model given varying variances across the data set. Note the vertical axis labels; $y(x)$ are the measurement outcomes, $m(x)$ is the mean function, i.e., the most likely model, $\rho(x)$ is the
    surrogate model, often assumed to equal the mean function and $f(x)$ is
    the "ground truth" latent function.}
    \label{fig:non_iid}
\end{figure}

%% file: figures/figure4.tex
\begin{figure}[htbp]
    \centering
    \includegraphics[
    width = \linewidth]{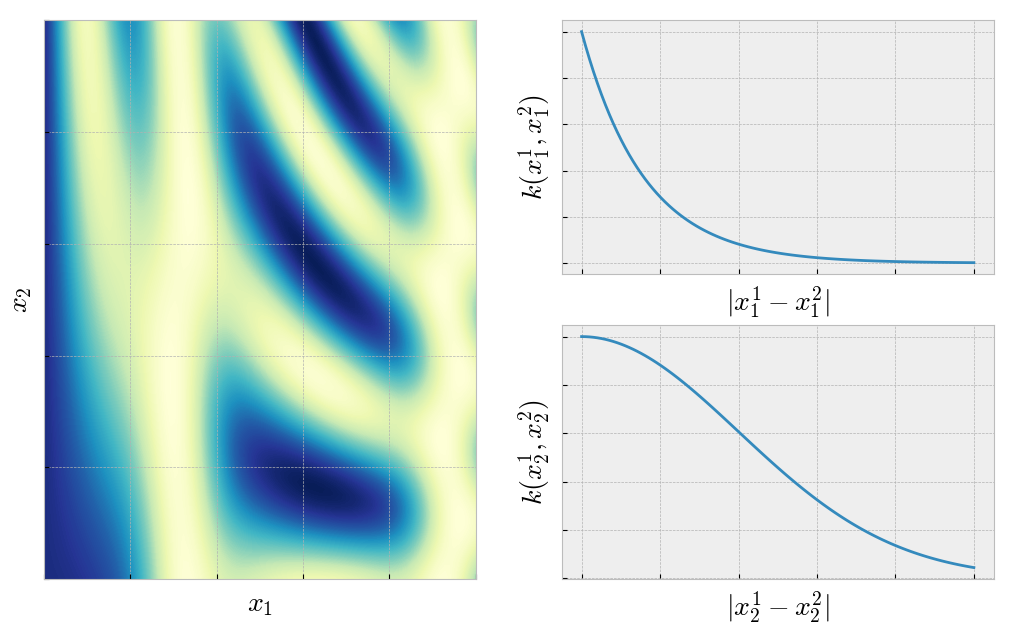}
    \caption{Model function with 
    different length scales and different orders of 
    differentiability in different directions.
    In $x_1$ direction we have assumed that the model function is not differentiable. Therefore we used
    the exponential kernel. In $x_2$
    direction, the model can be differentiated an infinite number of times. We therefore chose the squared exponential kernel.
    For other orders of differentiability, other kernels can be used. Fixing the order of differentiability also
    gives the user the ability to incorporate domain knowledge into the experiment.}
    \label{fig:anisotropic_kernels}
\end{figure}

%% file: figures/figure5.tex
\begin{figure}[!ht]
  \centering
   \input{./figures/3d.pdf_tex}
   \caption{The result of the diffusion-coefficient example on a three-dimensional input space. 
   The figure shows the result of the GP approximation
   after 500 measurements for three different nanoparticle radii. While the measurement results are always subject to differing noise, the model
   can take noise into account in different ways. Most commonly noise is ignored (left column). If noise is included, it is common
   to approximate it by i.i.d.~noise (middle column). The proposed method models the noise as what it is, which is non-i.i.d.~noise (right column).
   The iso-lines of the approximation are shown in white while the iso-lines of the ground truth are shown in red. Observe how
   the no-noise and the i.i.d.~noise approximations create localized artifacts. The non-i.i.d.~approximation does a far better job of
   creating a smooth model that explains all data including noise.}
   \label{fig:3dnoise}
 \end{figure}
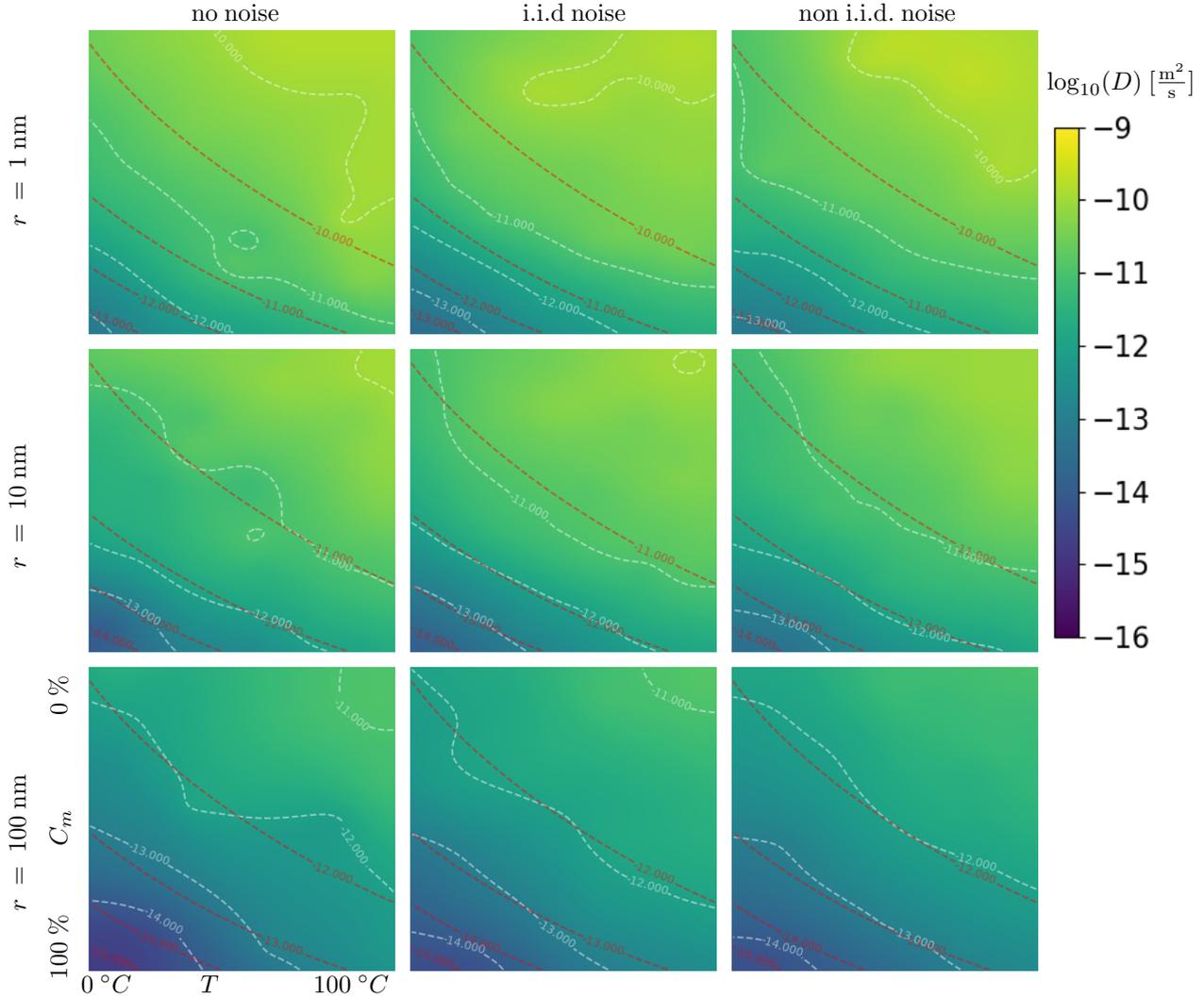

%% file: figures/3d.pdf_tex
\begingroup%
  \makeatletter%
  \providecommand\color[2][]{%
    \errmessage{(Inkscape) Color is used for the text in Inkscape, but the package 'color.sty' is not loaded}%
    \renewcommand\color[2][]{}%
  }%
  \providecommand\transparent[1]{%
    \errmessage{(Inkscape) Transparency is used (non-zero) for the text in Inkscape, but the package 'transparent.sty' is not loaded}%
    \renewcommand\transparent[1]{}%
  }%
  \providecommand\rotatebox[2]{#2}%
  \ifx\svgwidth\undefined%
    \setlength{\unitlength}{545.52729492bp}%
    \ifx\svgscale\undefined%
      \relax%
    \else%
      \setlength{\unitlength}{\unitlength * \real{\svgscale}}%
    \fi%
  \else%
    \setlength{\unitlength}{\svgwidth}%
  \fi%
  \global\let\svgwidth\undefined%
  \global\let\svgscale\undefined%
  \makeatother%
  \begin{picture}(1,0.65344618)%
    \put(0,0){\includegraphics[width = \linewidth]{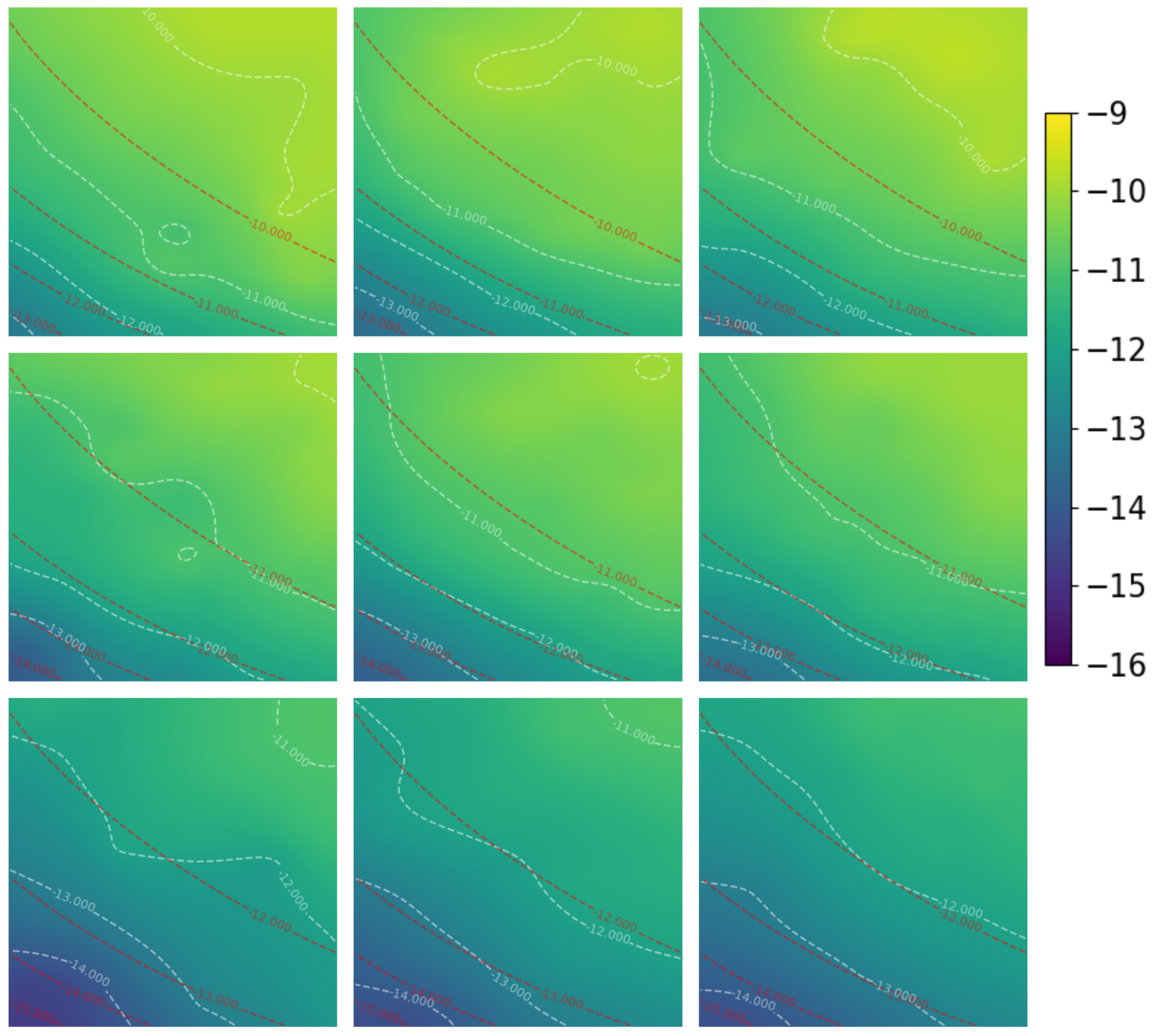}}
    \put(0.52,0.7){\color[rgb]{0,0,0}\rotatebox{0}{\makebox(0,0)[lb]{\smash{$\mathrm{non~i.i.d.~noise}$}}}}
    \put(0.32,0.7){\color[rgb]{0,0,0}\rotatebox{0}{\makebox(0,0)[lb]{\smash{$\mathrm{i.i.d~noise}$}}}}
    \put(0.08,0.7){\color[rgb]{0,0,0}\rotatebox{0}{\makebox(0,0)[lb]{\smash{$\mathrm{no~noise}$}}}}
    \put(-0.04,0.55){\color[rgb]{0,0,0}\rotatebox{90}{\makebox(0,0)[lb]{\smash{$r~=~1~\mathrm{nm}$}}}}
    \put(-0.04,0.30){\color[rgb]{0,0,0}\rotatebox{90}{\makebox(0,0)[lb]{\smash{$r~=~10~\mathrm{nm}$}}}}
    \put(-0.04,0.05){\color[rgb]{0,0,0}\rotatebox{90}{\makebox(0,0)[lb]{\smash{$r~=~100~\mathrm{nm}$}}}}
    \put(-0.00,-0.01){\color[rgb]{0,0,0}\rotatebox{0}{\makebox(0,0)[lb]{\smash{{$0~^{\circ}C ~~~~~~~~~~~T ~~~~~~~~~~~~~~~100~^{\circ}C$}}}}}
    \put(-0.01,0.0){\color[rgb]{0,0,0}\rotatebox{90}{\makebox(0,0)[lb]{\smash{{$100~\% ~~~~~~~~~~~C_m ~~~~~~~~~~~~~~~0~\%$}}}}}
    \put(0.7,0.65){\color[rgb]{0,0,0}\rotatebox{0}{\makebox(0,0)[lb]{\smash{$\mathrm{log}_{10}(D)~[\frac{\mathrm{m}^2}{\mathrm{s}}]$}}}}
  \end{picture}%
\endgroup%


%% file: figures/figure6.tex
\begin{figure}[!ht]
    \centering
    \includegraphics[width = \linewidth]{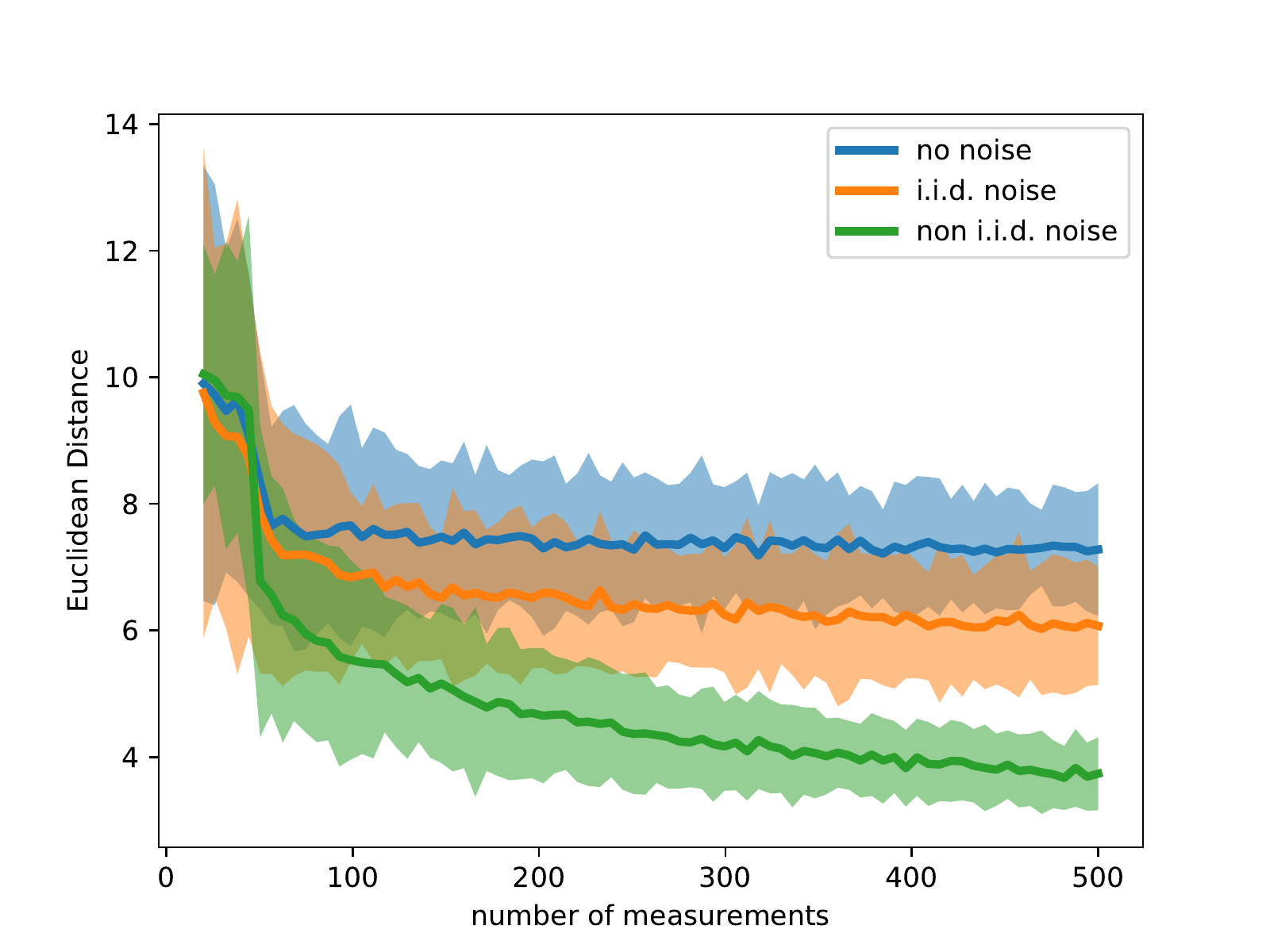}
    \caption{The approximation errors of the surrogate model during the diffusion-coefficient example 
    (Figure \ref{fig:3dnoise}), for three different noise models noted in the legend. The bands around each line represent the standard deviation of this error metric computed by running repeated synthetic experiments.}
    \label{fig:3derror}
\end{figure}

%% file: figures/figure7.tex
\begin{figure}[!ht]
    \centering
    \begin{subfigure}[t]{\linewidth}
    \includegraphics[trim={3.8cm 0cm 7.5cm 0cm},clip,width = 16.5cm]{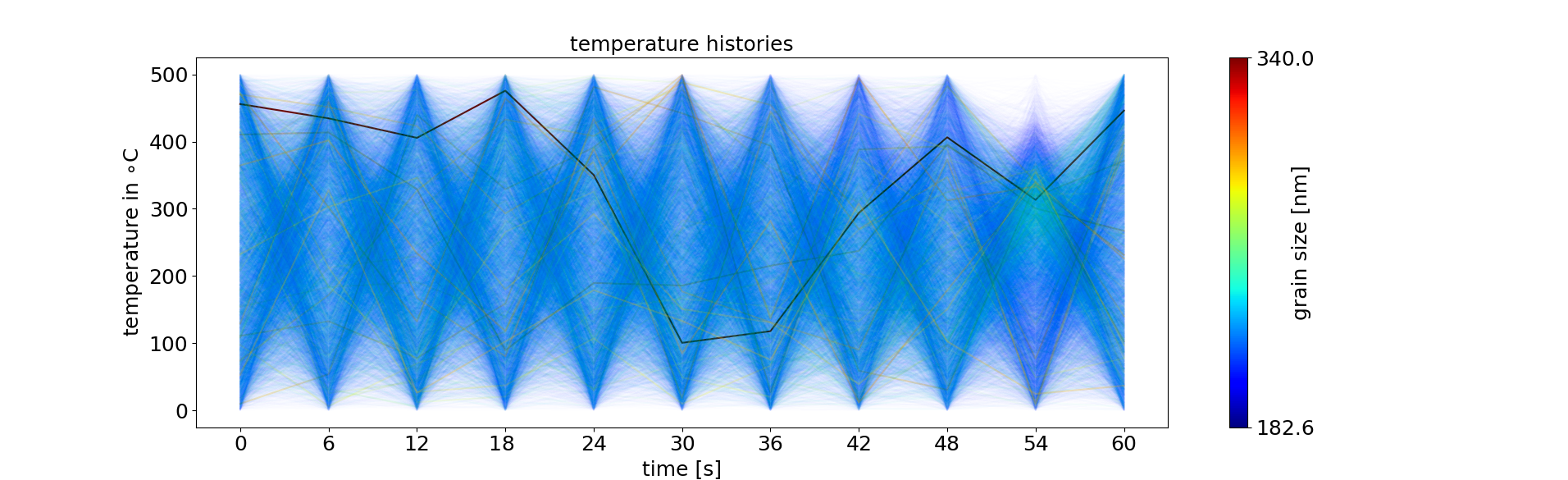}
    \caption{}
    \end{subfigure}
    \hspace*{-0.4cm}
    \begin{subfigure}[t]{0.41\linewidth}
    \includegraphics[trim={0.0cm 0cm 3.6cm 0cm},clip,height = 7cm]{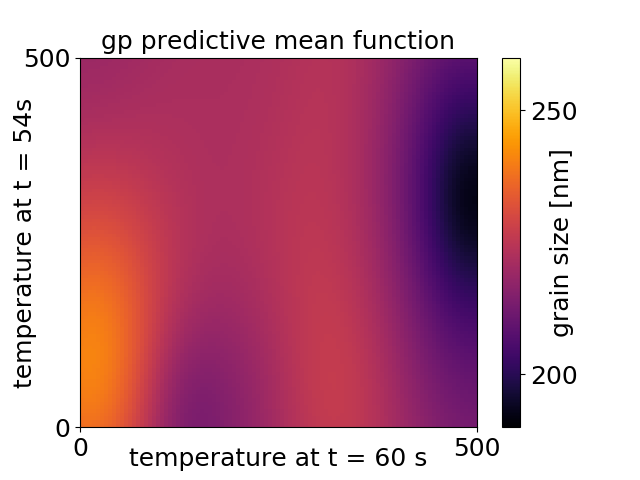}
    \caption{}
    \end{subfigure}
    ~~~~~~~~~~~~~~
    \begin{subfigure}[t]{0.50\linewidth}
    \includegraphics[trim={0.0cm 0cm 1.2cm 0cm},clip,height = 7 cm]{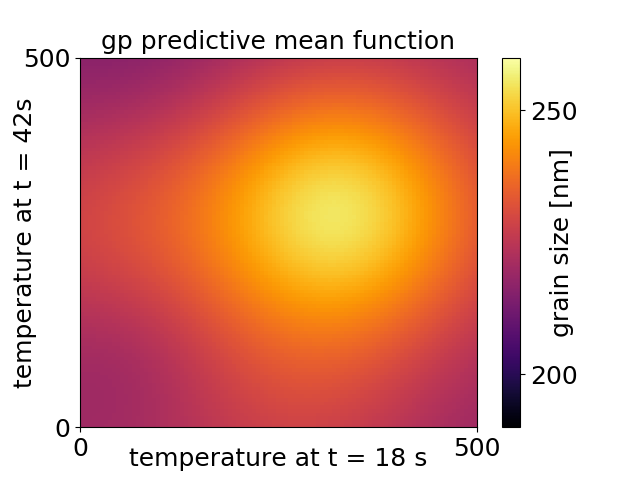}
    \caption{}
    \end{subfigure}
    
    \hspace*{-2.5cm}
    \begin{subfigure}[t]{\linewidth}
    \includegraphics[trim={0.0cm 0cm 3cm 1cm},clip,width = 16.2cm]{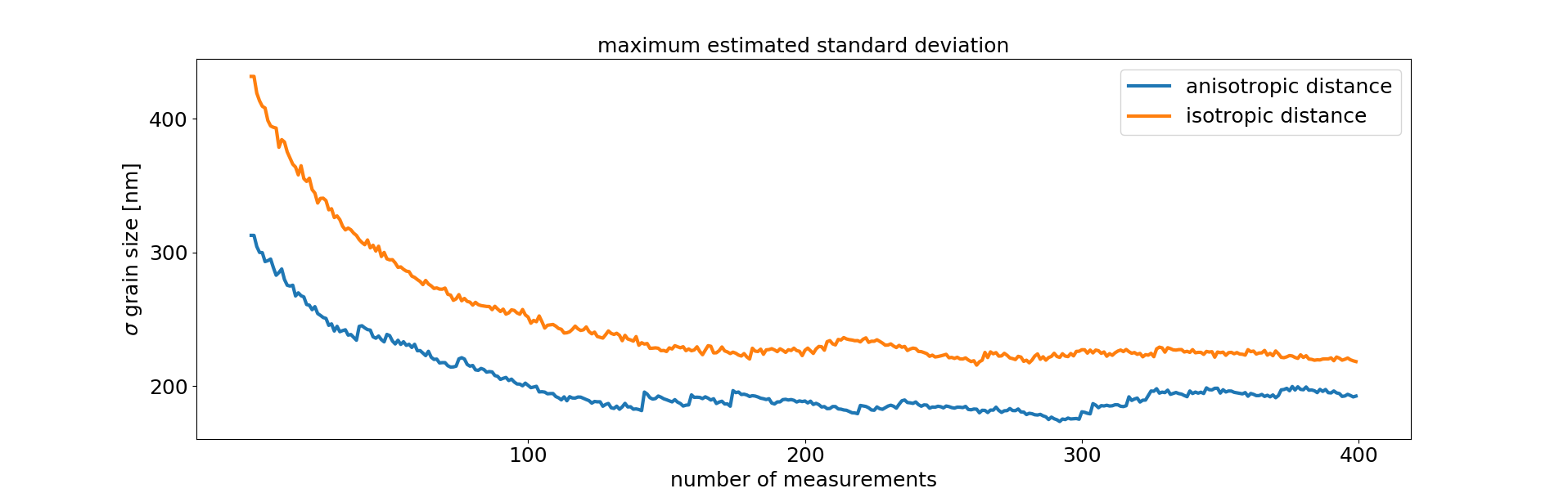}
    \caption{}
    \end{subfigure}
    \caption{Visualization of the grain size as a function of temperature history for a simple model of block copolymer grain size
    coarsening. 
    The figure demonstrates that when describing physical systems in high-dimensional spaces, strong anisotropy is frequently observed; only by taking this into account when estimating errors, will experimental guidance be optimal. 
    (a) $10,000$ simulated temperature histories and their corresponding grain size represented by color. The majority of histories terminate in a small grain size (blue lines). A small select set of histories yield large grain sizes (dark red lines). 
    (b) Example two-dimensional slice through the 11-dimensional parameter space. The anisotropy is clearly visible. (c) A different two-dimensional slice with no significant anisotropy present. (d) The estimated
    maximum standard deviation across the 11-dimensional domain as function of the number of measurements during a synthetic autonomous experiment.}
    \label{fig:11d}
\end{figure}

%% file: figures/figure8.tex
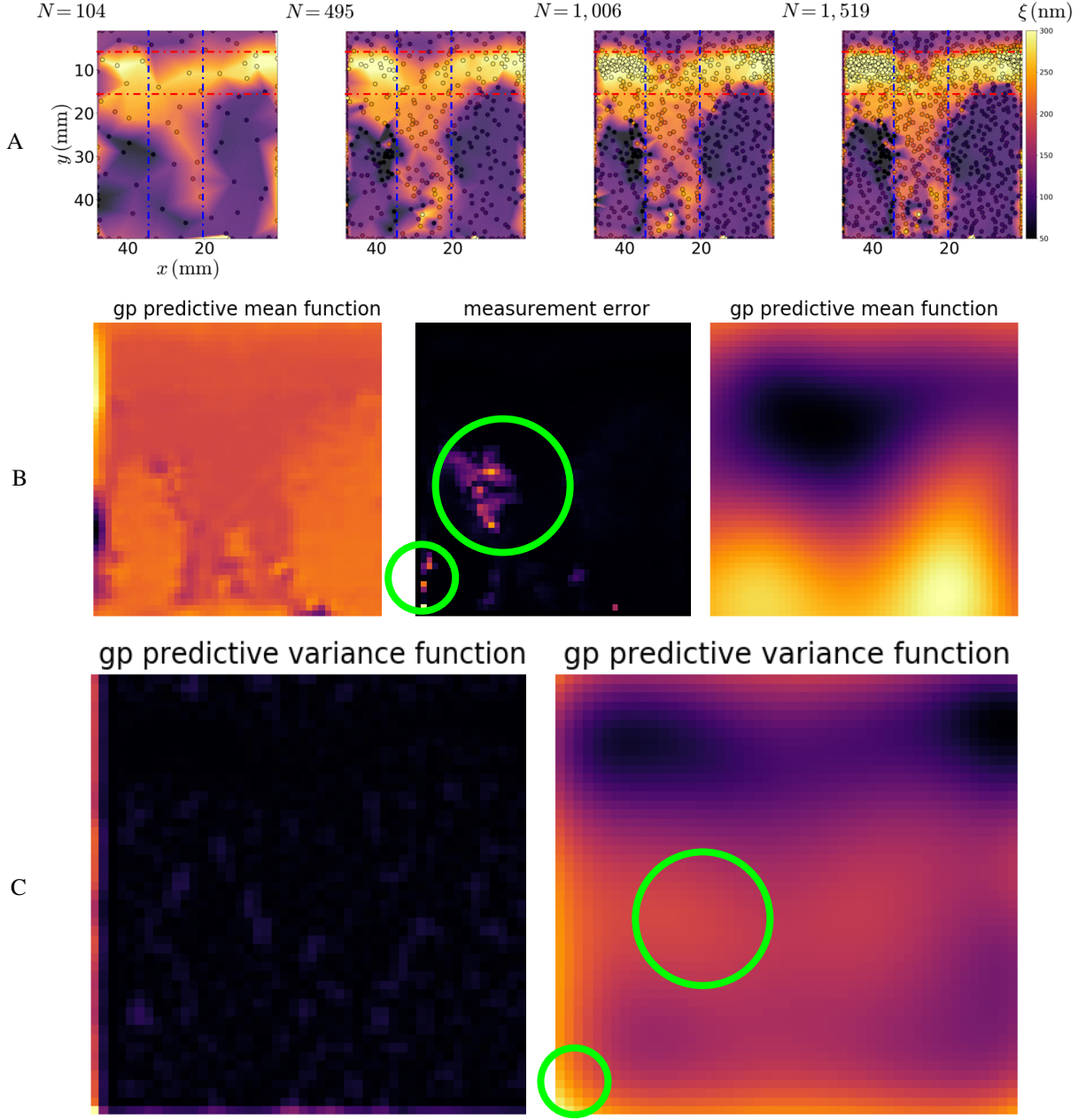
\begin{figure}[!htbp]
    \centering
    \begin{subfigure}[t]{0.24\linewidth}
    \includegraphics[width = \linewidth]{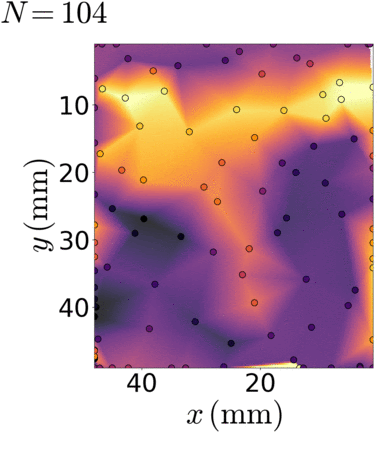}
    \end{subfigure}
    \begin{subfigure}[t]{0.24\linewidth}
    \includegraphics[width = \linewidth]{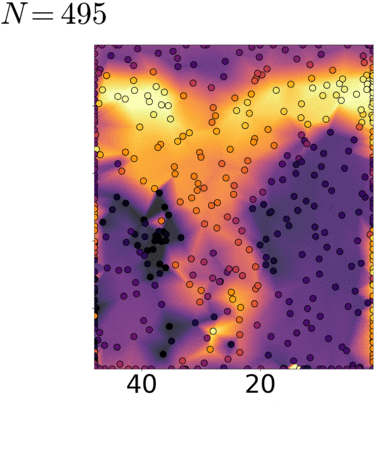}
    \end{subfigure}
    \begin{subfigure}[t]{0.24\linewidth}
    \includegraphics[clip,width = \linewidth]{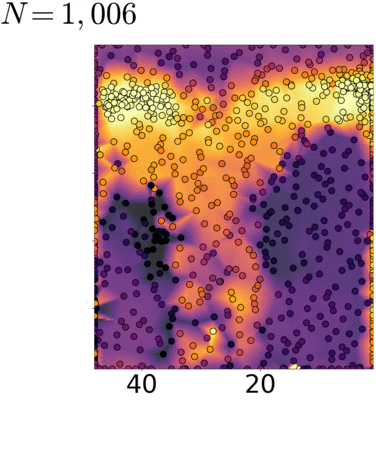}
    \end{subfigure}
    \begin{subfigure}[t]{0.24\linewidth}
    \includegraphics[width = 4.3cm]{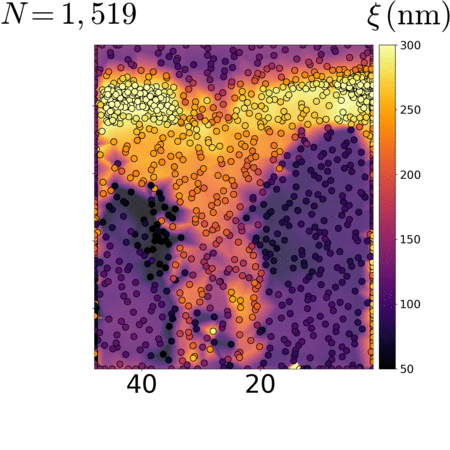}
    \end{subfigure}
    \put(-395,101){ \tikz [color = red, line width = 3pt] \draw[thick, dash dot] (0,1) -- (2.7,1);}
    \put(-395,83){ \tikz [color = red, line width = 3pt] \draw[thick, dash dot] (0,1) -- (2.7,1);}
    \put(-290,101){ \tikz [color = red, line width = 3pt] \draw[thick, dash dot] (0,1) -- (2.7,1);}
    \put(-290,83){ \tikz [color = red, line width = 3pt] \draw[thick, dash dot] (0,1) -- (2.7,1);}
    \put(-185,101){ \tikz [color = red, line width = 3pt] \draw[thick, dash dot] (0,1) -- (2.7,1);}
    \put(-185,83){ \tikz [color = red, line width = 3pt] \draw[thick, dash dot] (0,1) -- (2.7,1);}
    \put(-80,101){ \tikz [color = red, line width = 3pt] \draw[thick, dash dot] (0,1) -- (2.7,1);}
    \put(-80,83){ \tikz [color = red, line width = 3pt] \draw[thick, dash dot] (0,1) -- (2.7,1);}
    \put(-373,22){ \tikz [color = blue, line width = 3pt] \draw[thick, dash dot] (1,0) -- (1,3.1);}    
    \put(-350,22){ \tikz [color = blue, line width = 3pt] \draw[thick, dash dot] (1,0) -- (1,3.1);}
    \put(-268,22){ \tikz [color = blue, line width = 3pt] \draw[thick, dash dot] (1,0) -- (1,3.1);}    
    \put(-245,22){ \tikz [color = blue, line width = 3pt] \draw[thick, dash dot] (1,0) -- (1,3.1);}
    \put(-163,22){ \tikz [color = blue, line width = 3pt] \draw[thick, dash dot] (1,0) -- (1,3.1);}    
    \put(-140,22){ \tikz [color = blue, line width = 3pt] \draw[thick, dash dot] (1,0) -- (1,3.1);}
    \put(-58,22){ \tikz [color = blue, line width = 3pt] \draw[thick, dash dot] (1,0) -- (1,3.1);}    
    \put(-35,22){ \tikz [color = blue, line width = 3pt] \draw[thick, dash dot] (1,0) -- (1,3.1);}
    \put(-430,60){A}%
    
    \hspace*{0.0cm}
    \begin{subfigure}[t]{0.30\linewidth}
    \includegraphics[trim={1cm 1cm 1cm 0.5cm},clip,width = 5cm, height = 5cm]{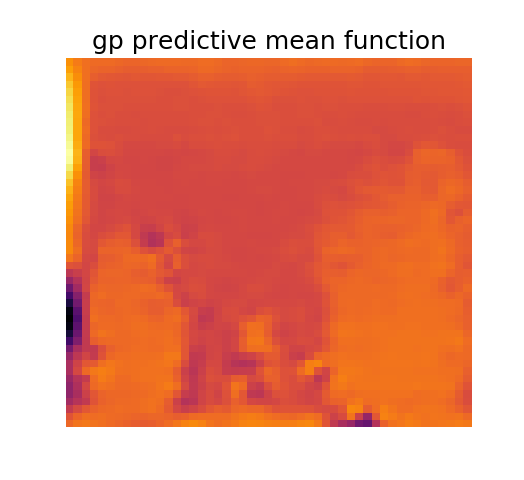}
    \end{subfigure}
    \begin{subfigure}[t]{0.30\linewidth}
    \includegraphics[trim={2.5cm 1cm 2.5cm 0.5cm},clip,width = 5cm, height = 5cm]{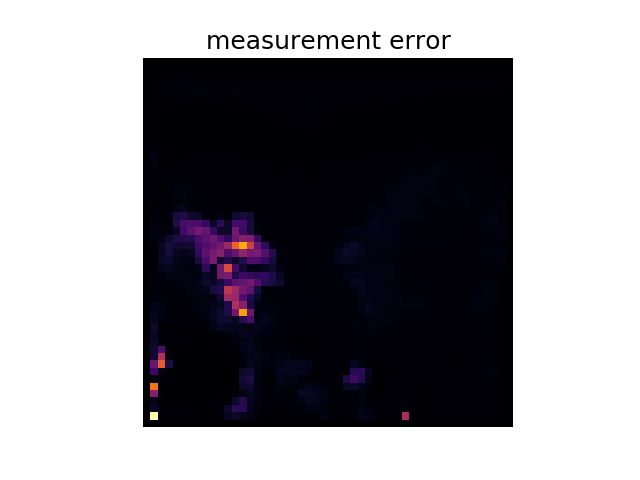}
    \end{subfigure}
    \begin{subfigure}[t]{0.30\linewidth}
    \includegraphics[trim={1cm 1cm 1cm 0.5cm},clip,width = 5cm, height = 5cm]{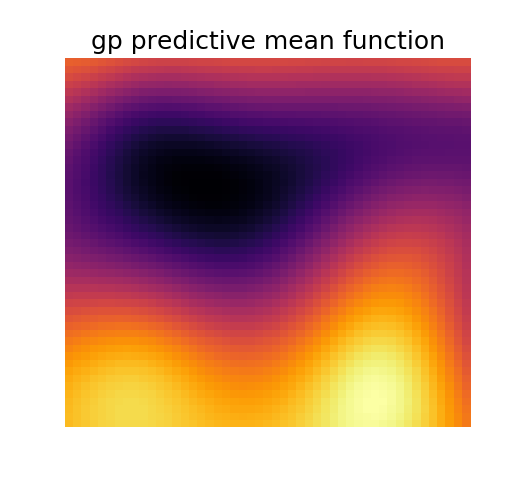}
    \end{subfigure}
    \put(-240,30){ \tikz [color = green, line width = 3pt] \draw (0,0) ellipse (1cm and 1cm);}%
    \put(-260,5){ \tikz [color = green, line width = 3pt] \draw (0,0) ellipse (0.5cm and 0.5cm);}%
    \put(-415,60){B}%
    
    \hspace*{-0.3cm}
    \begin{subfigure}[t]{0.45\linewidth}
    \includegraphics[trim={1.1cm 1cm 0.5cm 0.5cm},clip,width = 7.5cm, height = 7.5cm]{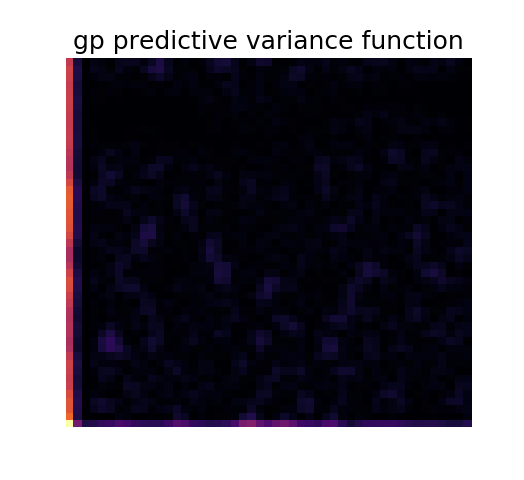}
    \end{subfigure}
    \begin{subfigure}[t]{0.45\linewidth}
    \includegraphics[trim={1cm 1cm 1cm 0.5cm},clip,width = 7.5cm, height = 7.5cm]{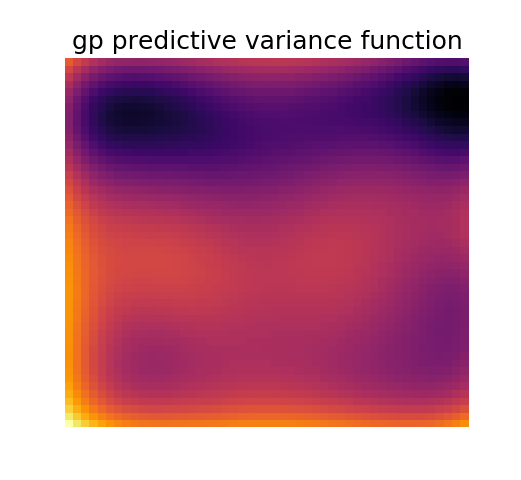}
    \end{subfigure}
    \put(-150,60){ \tikz [color = green, line width = 3pt] \draw (0,0) ellipse (1cm and 1cm);}%
    \put(-190,5){ \tikz [color = green, line width = 3pt] \draw (0,0) ellipse (0.5cm and 0.5cm);}%
    \put(-410,100){C}%
    \caption{(top row, A) Results of an autonomous SAXS experiment probing the distribution of grain size ($\xi$) in a combinatorial nanocomposite sample, as a function of coordinates $(x,\,y)$ representing a two-dimensional sample-processing parameter space, for increasing number of measurements ($N$). The sample consisted of a flow-coated film of polymer-grafted nano-rods on a surface-treated substrate, where the substrate surface energy increased linearly from $30.5 \, \mathrm{mN/m}$ (hydrophobic) at $x = 0$ to $70.2 \, \mathrm{mN/m}$ (hydrophilic) at $x \approx 50 \, \mathrm{mm}$, and the coating speed increased at constant acceleration ($0.002 \, \mathrm{mm/s^{2}}$) from $0 \, \mathrm{mm/s}$ (thicker film) at $y = 0$ to $0.45 \, \mathrm{mm/s}$ (thinner film) at $y \approx 50 \, \mathrm{mm}$. The autonomous experiment successfully identified a well ordered region (between red lines) that corresponded to uniform monolayer domains. Blue lines mark the region of solution-meniscus instability (see text).
    The points show the locations
    of measured data points; the same axes and orientation are used in subsequent plots in this figure. 
    (middle, row B, from the left) An exact Gaussian-process interpolation
    of the complete measured data-set for the peak position $q$. 
    The data is corrupted by measurement errors which corrupt the model
    if standard, exact interpolation techniques are used (including GPR). 
    The green circles mark the regions of the largest variances in the model
    and the corresponding high errors (measurement variances) that were recorded during the experiment. 
    On the right is the Gaussian process model of $q$, taking into account the non-i.i.d.~measurement variances.
    This model does not show any of the artifacts that are 
    visible in the exact GPR interpolation. 
    (bottom row, C) The final objective functions for no noise and non-i.i.d.~noise in $q$ which has to
    be maximized to determine the next optimal measurement. If the experiment had been steered using the posterior variances 
    in $q$ without accounting for non-i.i.d.~observation noise, the autonomous experiments would have been misled significantly.}
    \label{fig:time_evo}
\end{figure}

%% file: main.bbl
\begin{thebibliography}{37}
\providecommand{\natexlab}[1]{#1}
\providecommand{\url}[1]{\texttt{#1}}
\expandafter\ifx\csname urlstyle\endcsname\relax
  \providecommand{\doi}[1]{doi: #1}\else
  \providecommand{\doi}{doi: \begingroup \urlstyle{rm}\Url}\fi

\bibitem[Almgren et~al.(2017)Almgren, DeMar, Vetter, Riley, Antypas, Bard,
  Coffey, Dart, Dosanjh, Gerber, et~al.]{almgren2017advanced}
Ann Almgren, Phil DeMar, Jeffrey Vetter, Katherine Riley, Katie Antypas,
  Deborah Bard, Richard Coffey, Eli Dart, Sudip Dosanjh, Richard Gerber, et~al.
\newblock Advanced scientific computing research exascale requirements review.
  an office of science review sponsored by advanced scientific computing
  research, september 27-29, 2016, rockville, maryland.
\newblock Technical report, Argonne National Lab.(ANL), Argonne, IL (United
  States). Argonne Leadership~…, 2017.

\bibitem[Balachandran et~al.(2016)Balachandran, Xue, Theiler, Hogden, and
  Lookman]{balachandran2016adaptive}
Prasanna~V Balachandran, Dezhen Xue, James Theiler, John Hogden, and Turab
  Lookman.
\newblock Adaptive strategies for materials design using uncertainties.
\newblock \emph{Scientific reports}, 6:\penalty0 19660, 2016.

\bibitem[Ballabio et~al.(2019)Ballabio, Lugato, Fern{\'a}ndez-Ugalde, Orgiazzi,
  Jones, Borrelli, Montanarella, and Panagos]{ballabio2019mapping}
Cristiano Ballabio, Emanuele Lugato, Oihane Fern{\'a}ndez-Ugalde, Alberto
  Orgiazzi, Arwyn Jones, Pasquale Borrelli, Luca Montanarella, and Panos
  Panagos.
\newblock Mapping lucas topsoil chemical properties at european scale using
  gaussian process regression.
\newblock \emph{Geoderma}, 355:\penalty0 113912, 2019.

\bibitem[Bao et~al.(2018)Bao, Shaw, Gu, Toney, and Bao]{bao2018meniscus}
Xiaodan Bao, Leo Shaw, Kevin Gu, Michael~F. Toney, and Zhenan Bao.
\newblock The meniscus-guided deposition of semiconducting polymers.
\newblock \emph{Nature Communications}, 9:\penalty0 534, 2018.

\bibitem[Bijl(2016)]{bijl2016gaussian}
H~Bijl.
\newblock Gaussian process regression techniques with applications to wind
  turbines.
\newblock \emph{Delft University of Technology, Doctoral degree}, 2016.

\bibitem[Cang et~al.(2018)Cang, Li, Yao, Jiao, and Ren]{cang2018improving}
Ruijin Cang, Hechao Li, Hope Yao, Yang Jiao, and Yi~Ren.
\newblock Improving direct physical properties prediction of heterogeneous
  materials from imaging data via convolutional neural network and a
  morphology-aware generative model.
\newblock \emph{Computational Materials Science}, 150:\penalty0 212--221, 2018.

\bibitem[Che et~al.(2016)Che, Park, Grabowski, Jawaid, Kelley, Koerner, and
  Vaia]{che2016preparation}
Justin Che, Kyoungweon Park, Christopher~A Grabowski, Ali Jawaid, John Kelley,
  Hilmar Koerner, and Richard~A Vaia.
\newblock Preparation of ordered monolayers of polymer grated nanoparticles:
  Impact of architecture, concentration, and substrate surface energy.
\newblock \emph{Macromolecules}, 49:\penalty0 1834--1847, 2016.

\bibitem[Cheng(2008)]{cheng2008formula}
Nian-Sheng Cheng.
\newblock Formula for the viscosity of a glycerol- water mixture.
\newblock \emph{Industrial \& engineering chemistry research}, 47\penalty0
  (9):\penalty0 3285--3288, 2008.

\bibitem[Dean(2000)]{dean2000design}
Edwin~B Dean.
\newblock Design of experiments, 2000.

\bibitem[Doerk and Yager(2017)]{doerk2017beyond}
Gregory~S Doerk and Kevin~G Yager.
\newblock Beyond native block copolymer morphologies.
\newblock \emph{Molecular Systems Design \& Engineering}, 2\penalty0
  (5):\penalty0 518--538, 2017.

\bibitem[Fisher(1992)]{fisher1992arrangement}
Ronald~A Fisher.
\newblock The arrangement of field experiments.
\newblock In \emph{Breakthroughs in statistics}, pages 82--91. Springer, 1992.

\bibitem[Forrester et~al.(2008)Forrester, Sobester, and
  Keane]{forrester2008engineering}
Alexander Forrester, Andras Sobester, and Andy Keane.
\newblock \emph{Engineering design via surrogate modelling: a practical guide}.
\newblock John Wiley \& Sons, 2008.

\bibitem[Genzer et~al.(2002)Genzer, Efimenko, and Fischer]{genzer2002molecular}
Jan Genzer, Kirill Efimenko, and Daniel~A Fischer.
\newblock Molecular orientation and grafting density in semifluorinated
  self-assembled monolayers of mono-, di-, and trichloro silanes on silica
  substrates.
\newblock \emph{Langmuir}, 18\penalty0 (24):\penalty0 9307--9311, 2002.

\bibitem[Gerber et~al.(2018)Gerber, Hack, Riley, Antypas, Coffey, Dart,
  Straatsma, Wells, Bard, Dosanjh, et~al.]{gerber2018crosscut}
Richard Gerber, James Hack, Katherine Riley, Katie Antypas, Richard Coffey, Eli
  Dart, Tjerk Straatsma, Jack Wells, Deborah Bard, Sudip Dosanjh, et~al.
\newblock Crosscut report: Exascale requirements reviews, march 9--10,
  2017--tysons corner, virginia. an office of science review sponsored by:
  Advanced scientific computing research, basic energy sciences, biological and
  environmental research, fusion energy sciences, high energy physics, nuclear
  physics.
\newblock Technical report, Oak Ridge National Lab.(ORNL), Oak Ridge, TN
  (United States); Argonne~…, 2018.

\bibitem[Godaliyadda et~al.(2016)Godaliyadda, Ye, Uchic, Groeber, Buzzard, and
  Bouman]{godaliyadda2016supervised}
GM~Godaliyadda, Dong~Hye Ye, Michael~D Uchic, Michael~A Groeber, Gregery~T
  Buzzard, and Charles~A Bouman.
\newblock A supervised learning approach for dynamic sampling.
\newblock \emph{Electronic Imaging}, 2016\penalty0 (19):\penalty0 1--8, 2016.

\bibitem[Habib et~al.(2016)Habib, Roser, Gerber, Antypas, Riley, Williams,
  Wells, Straatsma, Almgren, Amundson, et~al.]{habib2016ascr}
Salman Habib, Robert Roser, Richard Gerber, Katie Antypas, Katherine Riley, Tim
  Williams, Jack Wells, Tjerk Straatsma, A~Almgren, J~Amundson, et~al.
\newblock Ascr/hep exascale requirements review report.
\newblock \emph{arXiv preprint arXiv:1603.09303}, 2016.

\bibitem[Hanuka et~al.(2019)Hanuka, Duris, Shtalenkova, Kennedy, Edelen,
  Ratner, and Huang]{hanuka2019online}
A~Hanuka, J~Duris, J~Shtalenkova, D~Kennedy, A~Edelen, D~Ratner, and X~Huang.
\newblock Online tuning and light source control using a physics-informed
  gaussian process adi.
\newblock \emph{arXiv preprint arXiv:1911.01538}, 2019.

\bibitem[Huang et~al.(2006)]{huang2006technical}
Tzu-Kuo Huang et~al.
\newblock A technical introduction to gaussian process regression.
\newblock 2006.

\bibitem[Jain et~al.(2013)Jain, Ong, Hautier, Chen, Richards, Dacek, Cholia,
  Gunter, Skinner, Ceder, et~al.]{jain2013commentary}
Anubhav Jain, Shyue~Ping Ong, Geoffroy Hautier, Wei Chen, William~Davidson
  Richards, Stephen Dacek, Shreyas Cholia, Dan Gunter, David Skinner, Gerbrand
  Ceder, et~al.
\newblock Commentary: The materials project: A materials genome approach to
  accelerating materials innovation.
\newblock \emph{Apl Materials}, 1\penalty0 (1):\penalty0 011002, 2013.

\bibitem[Kuss(2006)]{kuss2006gaussian}
Malte Kuss.
\newblock \emph{Gaussian process models for robust regression, classification,
  and reinforcement learning}.
\newblock PhD thesis, Technische Universit{\"a}t, 2006.

\bibitem[Laboratory(2015{\natexlab{a}})]{SciAnalysis}
Brookhaven~National Laboratory.
\newblock Scianalysis.
\newblock \url{https://github.com/CFN-softbio/SciAnalysis}, 2015{\natexlab{a}}.

\bibitem[Laboratory(2015{\natexlab{b}})]{bluesky}
Brookhaven~National Laboratory.
\newblock Bluesky.
\newblock \url{https://github.com/NSLS-II/bluesky}, 2015{\natexlab{b}}.

\bibitem[Majewski and Yager(2016)]{majewski2016rapid}
Pawel~W Majewski and Kevin~G Yager.
\newblock Rapid ordering of block copolymer thin films.
\newblock \emph{Journal of Physics: Condensed Matter}, 28\penalty0
  (40):\penalty0 403002, 2016.

\bibitem[Mart{\'\i}nez et~al.(2007)Mart{\'\i}nez, Mart{\'\i}nez,
  P{\'e}rez-Ros{\'e}s, and Quir{\'o}s]{martinez2007image}
Alondra Mart{\'\i}nez, Jennifer Mart{\'\i}nez, Hebert P{\'e}rez-Ros{\'e}s, and
  Ricardo Quir{\'o}s.
\newblock Image processing using voronoi diagrams.
\newblock In \emph{IPCV}, pages 485--491, 2007.

\bibitem[McHutchon and Rasmussen(2011)]{mchutchon2011gaussian}
Andrew McHutchon and Carl~E Rasmussen.
\newblock Gaussian process training with input noise.
\newblock In \emph{Advances in Neural Information Processing Systems}, pages
  1341--1349, 2011.

\bibitem[McKay et~al.(1979)McKay, Beckman, and Conover]{mckay1979comparison}
Michael~D McKay, Richard~J Beckman, and William~J Conover.
\newblock Comparison of three methods for selecting values of input variables
  in the analysis of output from a computer code.
\newblock \emph{Technometrics}, 21\penalty0 (2):\penalty0 239--245, 1979.

\bibitem[Noack et~al.(2019)Noack, Yager, Fukuto, Doerk, Li, and
  Sethian]{noack2019kriging}
Marcus~M Noack, Kevin~G Yager, Masafumi Fukuto, Gregory~S Doerk, Ruipeng Li,
  and James~A Sethian.
\newblock A kriging-based approach to autonomous experimentation with
  applications to x-ray scattering.
\newblock \emph{Scientific Reports}, 9:\penalty0 11809, 2019.

\bibitem[Noack et~al.(2020)Noack, Doerk, Li, Fukuto, and
  Yager]{noack2020advances}
Marcus~M Noack, Gregory~S Doerk, Ruipeng Li, Masafumi Fukuto, and Kevin~G
  Yager.
\newblock Advances in kriging-based autonomous x-ray scattering experiments.
\newblock \emph{Scientific Reports}, 10:\penalty0 1325, 2020.

\bibitem[Pilania et~al.(2013)Pilania, Wang, Jiang, Rajasekaran, and
  Ramprasad]{pilania2013accelerating}
Ghanshyam Pilania, Chenchen Wang, Xun Jiang, Sanguthevar Rajasekaran, and
  Ramamurthy Ramprasad.
\newblock Accelerating materials property predictions using machine learning.
\newblock \emph{Scientific reports}, 3:\penalty0 2810, 2013.

\bibitem[Ruland and Smarsly(2004)]{ruland2004saxs}
Wilhelm Ruland and Bernd Smarsly.
\newblock Saxs of self-assembled oriented lamellar nano- composite films: an
  advanced method of evaluation.
\newblock \emph{Journal of Applied Crystallography}, 37:\penalty0 575--584,
  2004.

\bibitem[Santner et~al.(2003)Santner, Williams, Notz, and
  Williams]{santner2003design}
Thomas~J Santner, Brian~J Williams, William Notz, and Brain~J Williams.
\newblock \emph{The design and analysis of computer experiments}, volume~1.
\newblock Springer, 2003.

\bibitem[Scarborough et~al.(2017)Scarborough, Godaliyadda, Ye, Kissick, Zhang,
  Newman, Sheedlo, Chowdhury, Fischetti, Das, et~al.]{scarborough2017dynamic}
Nicole~M Scarborough, GM~Dilshan~P Godaliyadda, Dong~Hye Ye, David~J Kissick,
  Shijie Zhang, Justin~A Newman, Michael~J Sheedlo, Azhad~U Chowdhury, Robert~F
  Fischetti, Chittaranjan Das, et~al.
\newblock Dynamic x-ray diffraction sampling for protein crystal positioning.
\newblock \emph{Journal of synchrotron radiation}, 24\penalty0 (1):\penalty0
  188--195, 2017.

\bibitem[Schulz et~al.(2017)Schulz, Speekenbrink, and
  Krause]{schulz2017tutorial}
Eric Schulz, Maarten Speekenbrink, and Andreas Krause.
\newblock A tutorial on gaussian process regression with a focus on
  exploration-exploitation scenarios.
\newblock \emph{bioRxiv}, page 095190, 2017.

\bibitem[Stegle et~al.(2011)Stegle, Lippert, Mooij, Lawrence, and
  Borgwardt]{stegle2011efficient}
Oliver Stegle, Christoph Lippert, Joris~M Mooij, Neil~D Lawrence, and Karsten
  Borgwardt.
\newblock Efficient inference in matrix-variate gaussian models
  with$\backslash$iid observation noise.
\newblock In \emph{Advances in neural information processing systems}, pages
  630--638, 2011.

\bibitem[Thayer et~al.(2019)Thayer, Damiani, Dubrovin, Ford, Kroeger,
  O’Grady, Perazzo, Shankar, Weaver, Weninger, et~al.]{thayer2019data}
Jana Thayer, Daniel Damiani, Mikhail Dubrovin, Christopher Ford, Wilko Kroeger,
  Christopher~Paul O’Grady, Amedeo Perazzo, Murali Shankar, Matt Weaver,
  Clemens Weninger, et~al.
\newblock Data processing at the linac coherent light source.
\newblock In \emph{2019 IEEE/ACM 1st Annual Workshop on Large-scale
  Experiment-in-the-Loop Computing (XLOOP)}, pages 32--37. IEEE, 2019.

\bibitem[Vivarelli and Williams(1999)]{vivarelli1999discovering}
Francesco Vivarelli and Christopher~KI Williams.
\newblock Discovering hidden features with gaussian processes regression.
\newblock In \emph{Advances in Neural Information Processing Systems}, pages
  613--619, 1999.

\bibitem[Williams and Rasmussen(2006)]{williams2006gaussian}
Christopher~KI Williams and Carl~Edward Rasmussen.
\newblock \emph{Gaussian processes for machine learning}, volume~2.
\newblock MIT press Cambridge, MA, 2006.

\end{thebibliography}
